\pdfoutput=1
\documentclass[%
  article,notitlepage,
preprint,
 amsmath,amssymb,
 aps,
longbibliography,
 floats,
superscriptaddress]{revtex4-1}

\usepackage[version=3]{mhchem}
\usepackage[left=1.5cm, right=1.5cm, top=1.785cm, bottom=2.0cm]{geometry}
\usepackage{balance}
\usepackage{times,mathptmx}
\usepackage{graphicx} 
\usepackage{lastpage}
\usepackage[format=plain,justification=raggedright,singlelinecheck=false,font={stretch=1.125,small,sf},labelfont=bf,labelsep=space]{caption}
\usepackage{fancyhdr}
\usepackage{fnpos}
\usepackage[english]{babel}
\usepackage{array}
\usepackage{droidsans}
\usepackage{charter}
\usepackage[T1]{fontenc}
\usepackage[usenames,dvipsnames]{xcolor}
\usepackage{setspace}
\usepackage[compact]{titlesec}
\usepackage{fancyhdr}
\usepackage{subfig}
\usepackage{latexsym}
\usepackage{amsmath}
\usepackage{amsfonts}
\usepackage{amssymb}
\usepackage{bm}
\usepackage{color}

\usepackage[normalem]{ulem} 

\def \lp {\left(}
\def \rp {\right)}
\def \bv {\bm{v}}

\def \bpn {\bar{p}_n}
\def \bfn {\bar{\phi}_n}
\def \alco {\alpha^{\mathrm{co}}}
\def \Ca {\mathrm{Ca}}

\def \rc {\textcolor{blue}}

\usepackage{epstopdf}

\definecolor{cream}{RGB}{222,217,201}

\def \rc {\textcolor{blue}}

\def \mp {\mathcal{P}} 
\def \mpu {\mp_{0}} 

\def \In {^{\left(1\right)}}
\def \ot {^{\left(2\right)}}
\def \i  {^{\left(i\right)}}
\def \Uu {U_{0}}
\def \Ud {U_D}
\def \Us  {U_S}
\def \Uco {U_{SD}^{\mathrm{co}}}

\usepackage{epstopdf}

\definecolor{cream}{RGB}{222,217,201}

\begin{document}

\title{Swimming with a cage: Low-Reynolds-number locomotion 
inside a droplet}

\author{Shang Yik Reigh}
\email{reigh@is.mpg.de}
\affiliation{Department of Applied Mathematics and Theoretical Physics, 
  Center for Mathematical Science, University of Cambridge, Wilberforce Road,
  Cambridge CB3 0WA, United Kingdom}
\affiliation{
  Max-Plank-Institut f{\"u}r Intelligente Systeme,
  Heisenbergstra{\ss}e 3, 70569 Stuttgart, Germany}
\author{Lailai Zhu}
\email{lailai.zhu@epfl.ch}
\affiliation{
  Laboratory of Fluid Mechanics and Instabilities, 
  Ecole Polytechnique F{\'e}d{\'e}rale de Lausanne, Lausanne, CH-1015,
  Switzerland}
\affiliation{
  Current address: Linn\'{e} Flow Centre 
  and Swedish e-Science Research Centre (SeRC), KTH 
  Mechanics, SE-100 44 Stockholm, Sweden; Department of Mechanical and Aerospace Engineering,
  Princeton University, Princeton, NJ-08544, USA.}
\author{Fran{\c c}ois Gallaire}
\email{francois.gallaire@epfl.ch}
\affiliation{
Laboratory of Fluid Mechanics and Instabilities, 
  Ecole Polytechnique F{\'e}d{\'e}rale de Lausanne, Lausanne, CH-1015, Switzerland}
\author{Eric Lauga}
\email{e.lauga@damtp.cam.ac.uk}
\thanks{S. Y. Reigh and L. Zhu contributed equally to this work.}
\affiliation{Department of Applied Mathematics and Theoretical Physics, 
Center for Mathematical Science, University of Cambridge, Wilberforce Road,
Cambridge CB3 0WA, United Kingdom}




\pacs{Valid PACS appear here}

\begin{abstract}
Inspired by   recent
experiments  using synthetic microswimmers to manipulate droplets,
we investigate the low-Reynolds-number locomotion of a model swimmer (a spherical squirmer) encapsulated inside a 
droplet of comparable size in another viscous fluid. Meditated solely by hydrodynamic interactions, the encaged swimmer is seen to be able to  
propel the droplet, and in some situations both remain in a stable co-swimming state. The problem is tackled using both  an exact analytical theory and a numerical implementation based on boundary element method, with a particular focus on the  kinematics of the co-moving swimmer and droplet in a 
concentric configuration, and we obtain excellent  quantitative
agreement between the two. The droplet always moves slower than a  swimmer which uses  
purely tangential surface actuation but when it uses a particular combination of tangential and normal 
actuations, the squirmer and droplet are able to attain a same velocity and  stay concentric for all times. We next employ numerical simulations to examine the stability of their concentric co-movement, and highlight several stability  scenarios
 depending on the particular gait adopted by the swimmer. Furthermore, we show that the droplet reverses the nature of the  far-field flow induced by the swimmer: a droplet cage turns a pusher  swimmer into a puller, and vice versa. 
 Our work sheds light on the potential development of  droplets as self-contained carriers of both chemical content and self-propelled devices for  controllable 
and precise drug deliveries.
\end{abstract}

\maketitle

\section{Introduction}
Droplets have recently been used as small, isolated, aqueous compartments to encapsulate, incubate and manipulate cells 
for biological assays \cite{he2005selective}. Such droplet-based cell 
encapsulation is commonly accomplished in microfluidic 
devices which are able to  precisely produce and manipulate microdroplets of 
adjustable sizes\cite{koster2008drop,chabert2008microfluidic}. Current microfluidic technology
allows a high-throughput and controllable analysis to be performed on individual cells  
in their own discrete microenvironments. 

In related work, droplets have  been used to cage motile organisms such 
as the nematode {\it Caenorhabditis elegans} 
(\textit{C. elegans})\cite{clausell2008droplet, wen2015droplet} in order  to carry out developmental work.
In these studies, the size of an encaged adult \textit{C. elegans} is comparable to the droplet radius.
Despite their mobility, the worms failed to propel their liquid cages, because they were immobilized. In the work of Ref.~\cite{clausell2008droplet}, the droplet was tightly squeezed inside 
a capillary tube, forming a plug thus immobilized hydrodynamically by the lubrication film 
 while in the work of  
Ref.~\cite{wen2015droplet}, the droplet was anchored mechanically by a microfluidic trap.


Motivated by these droplet-based encapsulations of motile organisms, we raise in this paper a simple  question: is it possible for a  microswimmer 
encaged in a droplet to propel its viscous cage and co-swim with it?
One could envision setups of this type of interest to  the drug delivery community using 
droplets as small self-contained units   propelled and steered by their internal synthetic swimmers. 

Recently, microrobots  propelled by a magnetically-rotated helical  appendage mimicking the flagella of bacteria such as  {\it Escherichia coli} (\textit{E.~coli})
were fabricated\cite{zhang2009artificial,tottori2012magnetic},  encapsulated 
and operated inside a water-in-oil droplet in microfluidic chips\cite{ding:16}. In this case, the  droplets were not mobile, presumably  for  two reasons: the swimmer was much smaller than 
the droplet and the droplet was  large compared to the height of the micro-fluidic chips so that
it was tightly squeezed and thus anchored hydrodynamically\cite{clausell2008droplet}. Excitingly,
the same group  managed however to use their microrobots to push a droplet of  comparable size from the exterior when the droplet was unbounded or loosely bounded.


In this paper, we  conduct a combined theoretical and numerical study of a three-dimensional ($3$D) 
model microswimmer encapsulated in a droplet in free space. The size of the swimmer is of the same order as the radius of the  droplet and we attempt to answer the following fundamental questions: Will the droplet co-swim with the swimmer? What is the swimming 
velocity of the droplet compared to that of the swimmer? How are the kinematics and energetics of the microswimmer 
affected by the confinement due to the presence of the droplet? How stable is the co-movement of the concentric 
pair of swimmer 
and droplet?

\section{Problem description}
\begin{figure}[t!]
  \centering
  \includegraphics[scale=1.2,angle=0]{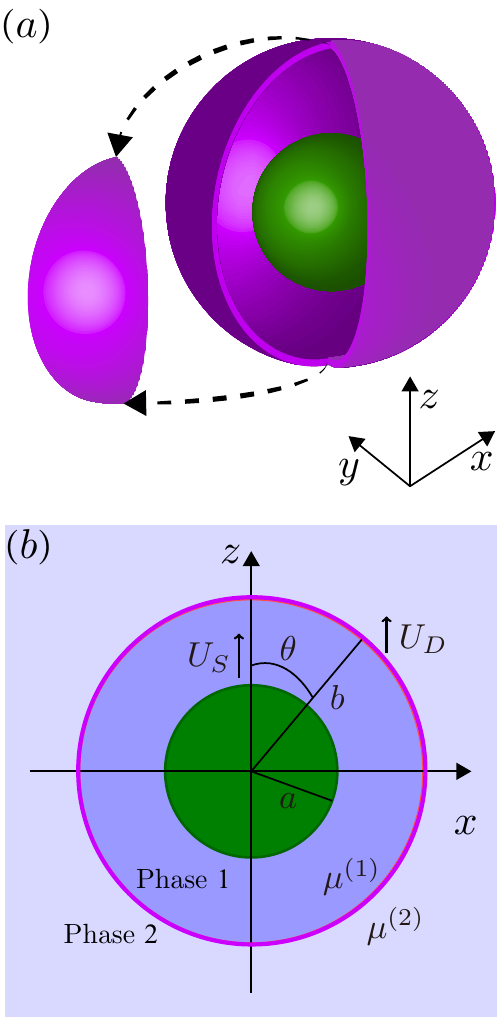}
  \caption{(a) Three-dimensional sketch of a spherical swimmer of radius $a$ (green) inside a spherical droplet of 
radius $b$ (magenta).
    (b) The squirmer and the droplet co-swim in the $z$ direction with a velocity of $\Us$ 
and $\Ud$, 
     respectively.  The fluids inside and outside 
     the droplet are marked as phase $1$ and phase $2$ and are distinguished by 
 their viscosity $\mu\In$ and $\mu\ot$, respectively.
  }
  \label{schem}
\end{figure}

We consider, in the creeping-flow  regime, the locomotion of a $3$D microswimmer encapsulated in a droplet. 
Due to hydrodynamic 
interactions, the motion of the swimmer 
is influenced by the presence of the droplet interface. The geometrical setup 
is shown in Fig.~\ref{schem}a.
We use a spherical, axisymmetric squirmer\cite{lighthill:52,blake:71} as our model swimmer. It achieves
locomotion by squirming, i.e.~by generating tangential and/or normal velocities on its fixed spherical surface. This is a classical  model for  physical actuation of  
microorganisms   continuously deforming their bodies
or beating their densely-packed cilia, and  has been employed in the past to address a variety of biophysical aspects 
of locomotion\cite{magar2003nutrient, 
ishikawa:06,michelin2010efficiency,doostmohammadi2012low,zottl2012nonlinear,pak:14,datt2015squirming, 
delfau2016collective}. 
The shape of the droplet is maintained  as spherical by maintaining  a sufficiently  large surface tension $\gamma$ on its 
interface, 
\textit{i.e.}~we assume to remain in the low-Capillary number limit. The radius
of the squirmer is denoted by $a$ while that of the droplet is $b > a$, respectively, and $\chi=b/a > 1$ is the size ratio.
The fluid phases inside and outside the droplet are marked as phase $1$ and $2$. Both are Newtonian, with dynamic 
viscosities  of ${\mu}\In$ and ${\mu}\ot$, and $\lambda=\mu\ot/\mu\In$ denotes 
the viscosity ratio.  
Both Cartesian ($x,y,z$) and spherical ($r,\theta,\phi$) coordinate 
systems are used, shown in Fig.~\ref{schem}b. 

We then solve the steady Stokes equations for fluid phase $1$ and $2$,
\begin{align}
  \nabla p\i = {\mu}^{\left( i \right)} \nabla^2 \bm{v}\i, \hspace{20pt} \nabla \cdot \bm{v}\i=0, \hspace{20pt}
  \label{stokes}
\end{align}
where  $p\i$ is the dynamic pressure and $\bm{v}\i$ the fluid 
velocity in phase $(i)$, where $i=1$ or $2$. 
Following classical work\cite{lighthill:52,blake:71}, we 
impose normal and/or tangential squirming velocities on the surface of the swimmer $r=a$ to represent its effective swimming
motion. These velocities are assumed to be time-independent and axisymmetric about its swimming direction, 
\textit{i.e.}, the $z$ axis  passing through the centers 
of the squirmer and droplet. The squirmer drives the droplet to co-swim in the 
same direction, and hence the problem is fully axisymmetric about the $z$ axis. The velocity of the swimmer and droplet 
are denoted by
$U_S$ and $U_D$ respectively.

In the laboratory frame of reference, the fluid velocity components  
$\bm{v}\In=(v_{r}\In,v_{\theta}\In)$ on the swimmer surface, $r=a$, are given by
\begin{align}
  v_{r}\In\vert_{r=a} &= \sum_{n=0}^{\infty}A_n P_n(\xi) + 
U_SP_1(\xi),\nonumber\\ 
  v_{\theta}\In\vert_{r=a} &= \sum_{n=1}^{\infty}B_n V_n(\xi) - U_SV_1(\xi),
  \label{bc1}
\end{align}
where $A_n$ (respectively $B_n$) indicates the $n$-th mode of the normal (respectively tangential) squirming velocities,
$P_n$ are the Legendre polynomial, $\xi \equiv \cos{\theta}$, $V_n=-2P_n^1(\xi)/(n^2+n)$, and $P_n^1$ is the associated 
Legendre function of the first kind of order 1. In Eq.~\eqref{bc1}, $U_S$ is the value of the unknown swimming velocity of the swimmer, and $U_D$ is the unknown swimming speed of the droplet. 

On the droplet interface $r=b$, the normal velocities in the droplet frame vanishes because
the droplet does not deform. In addition,  the tangential velocities and tangential stresses are continuous 
across the interface. These boundary conditions formulated in the laboratory frame are written as
\begin{align}
  &v_{r}\In\vert_{r=b} = v_{r}\ot\vert_{r=b}=U_D \cos{\theta},\nonumber \\
  &v_{\theta}\In\vert_{r=b} = v_{\theta}\ot\vert_{r=b},\nonumber \\ 
  &\Pi_{r\theta}\In\vert_{r=b}=\Pi\ot_{r\theta}\vert_{r=b},\label{bc2}
\end{align}
where $\boldsymbol{\Pi}\i = -p\i\mathbf{I}+\mu^{\left(i\right)} \left[\nabla\bm{v}\i + 
\left(\nabla\bm{v}\i\right)^{\mathrm{T}} \right] $ is the stress tensor for fluid $i$. Furthermore, the velocity $\bv\ot $ decays to zero in the far field $r \gg b$.

Finally, the total hydrodynamic forces exerted on both the swimmer and on the droplet interface are  zero, which  will be used to determine the values of both swimming velocities, $\Us$ and $\Ud$.
For an unbounded squirmer in a single-phase fluid,  the velocity $\Us\equiv \Uu$ 
is given by\cite{lighthill:52,blake:71}
\begin{align}\label{eq:us_blake}
\Uu = \frac{2B_1 - A_1}{3}\cdot
\end{align}

\section{Analytical theory}\label{sec:theory}

We first solve the problem analytically. The methodology is  based on Lamb's general solution of the Stokes equations in spherical coordinates\cite{lamb,happel:73}. 
For a single-phase fluid with viscosity $\mu$, the fluid velocity field $\bv$ can be expanded in  spherical harmonics as
\begin{align}
  \bm{v} = \sum_{n=-\infty}^{\infty} \Big[\nabla \phi_n 
  + \frac{n+3}{2\mu(n+1)(2n+3)}r^2\nabla p_n 
  -\frac{n}{\mu(n+1)(2n+3)}\bm{r}p_n \Big],
  \label{gensol}
\end{align}
where $p_n$ and $\phi_n$ are solid spherical harmonics satisfying   $\nabla^2 p_n = 0$ and $\nabla^2 \phi_n = 0$, respectively.
In axisymmetric flow, $p_n$ and $\phi_n$ are expressed by a series of Legendre functions as
\begin{align}
  p_n(r,\xi) = \tilde{p}_nr^n P_n(\xi), \hspace{20pt}
  \phi_n(r,\xi) = \tilde{\phi}_nr^n P_n(\xi), \nonumber
\end{align}
where $\tilde{p}_n$ and $\tilde{\phi}_n$ are constants independent of $r$ and $\xi$.

The radial and tangential  velocity components $v_r$ and $v_{\theta}$ are then obtained as 
\begin{align}\label{eq:exp}
  &v_r = \sum_{n\ge 0}^{\infty}\Big[\bar{p}_nr^{n+1}+\bar{\phi}_nr^{n-1}
  +\bar{p}_{-(n+1)}r^{-n} 
  +\bar{\phi}_{-(n+1)}\frac{1}{r^{n+2}}\Big] P_n(\xi),\nonumber\\
  &v_{\theta} = \sum_{n\ge 1}^{\infty} \Big[-\frac{n+3}{2}\bar{p}_n r^{n+1}
  -\frac{n+1}{2}\bar{\phi}_n r^{n-1}
  +\frac{n-2}{2}\bar{p}_{-(n+1)} {r^{-n}}  
  + \frac{n}{2}\bar{\phi}_{-(n+1)} r^{-\lp n+2 \rp} \Big]V_n(\xi),
\end{align}
where
\begin{align}
  \hspace{20pt}\bar{p}_n = \frac{n}{2\mu(2n+3)}\tilde{p}_n,
  \hspace{20pt}\bar{\phi}_n = n\tilde{\phi}_n. \nonumber
\end{align}
Note that the solution for the flow in region 1 may contain all terms in the brackets of Eq.~\eqref{eq:exp} while those in region 2 only contain the last two terms due to the boundary condition at infinity.

Applying this framework to our case, we  use Eq.~\ref{eq:exp} for both the inner and outer
fluid, solving for the unknown constants
 $\bpn\i$, $\bfn\i$,  $\bar p^{(i)}_{-(n+1)}$ and  $\bar \phi^{(i)}_{-(n+1)}$ ($i=1,2$) using 
the boundary conditions, Eqs.~\ref{bc1}-\ref{bc2}, together with the condition at infinity. Taking the  $n=0, 1$ terms in the series expansion of Eq.~\eqref{eq:exp}
with the use of  Eqs.~\ref{bc1}-\ref{bc2} leads to the system for the inner fluid
\begin{eqnarray} 
  &\bar{p}_{-1}\In + \frac{1}{a^2}\bar{\phi}_{-1}\In = A_0, 
  \hspace{15pt} \bar{p}\In_{-1} + \frac{1}{b^2}\bar{\phi}\In_{-1} = 0,\nonumber 
\\ 
  &a^2\bar{p}\In_{1} + \bar{\phi}\In_{1} + \frac{1}{a}\bar{p}\In_{-2} 
+\frac{1}{a^3}\bar{\phi}\In_{-2} = A_1 + U_S, 
\nonumber\\ 
  &-2a^2\bar{p}\In_{1} - \bar{\phi}\In_{1} - \frac{1}{2a}\bar{p}\In_{-2} 
+\frac{1}{2a^3}\bar{\phi}\In_{-2} = B_1 - 
U_S,\nonumber\\
  &b^2\bar{p}\In_{1} + \bar{\phi}\In_{1} + \frac{1}{b}\bar{p}\In_{-2} 
+\frac{1}{b^3}\bar{\phi}\In_{-2} = 
U_D,\nonumber\\ 
  &(-2-\frac{1}{\lambda})b^2\bar{p}\In_{1} - \bar{\phi}\In_{1} - 
\frac{1}{2b}\bar{p}\In_{-2} 
  + (\frac{1}{2}-\frac{1}{\lambda})\frac{1}{b^3}\bar{\phi}\In_{-2} = 
-\frac{1}{2}U_D.
  \label{udc}
\end{eqnarray}
Hence, the constants $\bar{p}\In_n$ and $\bar{\phi}\In_n$ ($n=-2, -1 \textrm{\; 
and \;} 1$) 
are obtained explicitly  in terms of both $U_S$ and $U_D$. The constants in the outer fluid are then given by
\begin{align}
  \bar{p}_{-2}\ot = \frac{1}{b^2}\bar{\phi}_{-2}\ot + bU_D,
  \hspace{15pt} \bar{\phi}_{-2}\ot = \frac{1}{\lambda} 
\Big(b^5\bar{p}_1\In+\bar{\phi}_{-2}\In\Big),
\end{align}
and the condition at infinity leads trivially to 
$  \bar{p}_{-1}\ot=0$ and
$ \bar{\phi}_{-1}\ot=0$.

Applying the force-free condition for the swimmer, we have 
\begin{equation}
\bm{F}= \int_{\hat{S}} 
\bm{\Pi}\In \cdot \hat{\bm{r}}dS=-4 \pi \nabla\left[r^3p\In_{-2}\right]=0,
\end{equation}
 which leads to  $\bar{p}\In_{-2}=0$. Applying the same condition for the 
droplet, we obtain $\bar{p}\ot_{-2}=0$. Plugging the two constants
into Eq.~\ref{udc}, we  obtain the values of all underdetermined constants  together with
the velocity of the swimmer,  $U_S$,   and that  of the droplet, $U_D$, as,
\begin{align}
  U_S = \frac{\Xi_1\lambda + \Xi_2}{\Delta},
  \label{vel_sw}
\end{align}
and
\begin{align}
  U_D = \frac{10(A_1+B_1)\chi^2}{\Delta},
  \label{vel_drop}
\end{align}
where
\begin{align}\label{eq:delta}
  &\Xi_1  = 2(2B_1-A_1)\chi^5 -10(A_1+B_1)\chi^2 +6(2A_1+B_1),\nonumber\\
  &\Xi_2  = 3(2B_1-A_1)\chi^5 +10(A_1+B_1)\chi^2 -6(2A_1+B_1),\nonumber\\
  &\Delta  = 3\left[2(\chi^5-1)\lambda+3\chi^5+2\right].
\end{align}
Similarly  to case of an unbounded squirmer (see Eq.~\ref{eq:us_blake}), 
the swimming velocities $\Us$ and $\Ud$ are seen to be independent of the squirming modes $A_n$ or $B_n$ for $n \geq 2$, but depend only on $A_1$ and $B_1$.

In order to complete the calculation and charactarize the flow in both fluids, we need to calculate the values of the constants $\bar{p}_n$, $\bar{\phi}_n$,
$\bar p_{-(n+1)}$ and  $\bar \phi_{-(n+1)}$ 
 for $n \geq 2$ in the series expansion from Eq.~\eqref{eq:exp}.
The velocities inside and outside the droplet in the laboratory frame are then obtained to be 
\begin{align}
  &v_{r}\In = \frac{A_0}{\chi^2-1}\Big\{\chi^2\Big(\frac{a}{r}\Big)^2-1 
\Big\}P_0(\xi)
  + \frac{A_1+B_1}{\Delta}\Big\{6(\lambda-1)\Big(\frac{r}{a}\Big)^2-10(\lambda-1)\chi^2
  +2(2\lambda+3)\chi^5\Big(\frac{a}{r}\Big)^3\Big\}P_1(\xi) \nonumber\\
  &\hspace{25pt}+\sum_{n=2}^{\infty}\frac{1}{\Delta_n}
  \Big\{(N_1A_n+N_2B_n)\Big(\frac{r}{a}\Big)^{n+1}+(N_3A_n+N_4B_n)\Big(\frac{r}{a}\Big)^{n-1}
  +(N_5A_n+N_6B_n)\Big(\frac{a}{r}\Big)^{n} \nonumber\\
  &\hspace{90pt}
  +(N_7A_n+N_8B_n)\Big(\frac{a}{r}\Big)^{n+2}\Big\} P_n(\xi),\nonumber\\
  &v_{\theta}\In =-\frac{A_1+B_1}{\Delta}
  \Big\{12(\lambda-1)\Big(\frac{r}{a}\Big)^2 -10(\lambda-1)\chi^2
  -(2\lambda+3)\chi^5\Big(\frac{a}{r}\Big)^3\Big\}V_1(\xi) \nonumber\\
  &\hspace{25pt}+\sum_{n=2}^{\infty}\frac{1}{\Delta_n}\Big\{-\frac{n+3}{2}(N_1A_n+N_2B_n)\Big(\frac{r}{a}\Big)^{n+1}
  -\frac{n+1}{2}(N_3A_n+N_4B_n)\Big(\frac{r}{a}\Big)^{n-1} \nonumber\\
  &\hspace{90pt} 
  +\frac{n-2}{2}(N_5A_n+N_6B_n)\Big(\frac{a}{r}\Big)^{n} 
  +\frac{n}{2}(N_7A_n+N_8B_n)\Big(\frac{a}{r}\Big)^{n+2}\Big\}
  V_n(\xi),\nonumber\\
  &v_{r}\ot = \frac{10(A_1+B_1)\chi^5}{\Delta} \Big(\frac{a}{r}\Big)^3P_1(\xi)
  -\sum_{n=2}^{\infty}\frac{c_1A_n+c_2B_n}{\Delta_n}
  \Big\{\frac{1}{\chi^2}\Big(\frac{a}{r}\Big)^{n} - \Big(\frac{a}{r}\Big)^{n+2}\Big\}
  P_n(\xi),\nonumber\\
  &v_{\theta}\ot = \frac{5(A_1+B_1)\chi^5}{\Delta} \Big(\frac{a}{r}\Big)^3 V_1(\xi)
  -\sum_{n=2}^{\infty}\frac{c_1A_n+c_2B_n}{2\Delta_n}
  \Big\{\frac{n-2}{\chi^2}\Big(\frac{a}{r}\Big)^{n}  - n\Big(\frac{a}{r}\Big)^{n+2}\Big\}
  V_n(\xi),
  \label{vel_full}
\end{align}
where the values of all undefined constants are provided in  Appendix A.

We can finally  calculate the power consumption  of the squirmer, $\mp$, which is equal to rate of working done by the squirmer on the fluid,
\begin{align}
  \mp = - \int_{\hat{S}} \bv\In \cdot \bm{\Pi}\In \cdot \bm{n}_{\hat{S}} d{S},
  \label{pw_sw}
\end{align}
where $\bm{n}_{\hat{S}}$ denotes the normal vector on $\hat{S}$ pointing towards
the fluid. We obtain
\begin{align}
  &\frac{\mp}{4\pi {\mu}_1 a} = \bigg[ 2\frac{2\chi^2+1}{\chi^2-1}A_0^2
  +2(Z_1+Z_2)Z_3\frac{(A_1+B_1)^2}{\Delta^2} \nonumber\\
  &\hspace{0pt}
  +\sum_{n\ge 2}^{\infty}\frac{1}{(2n+1)\Delta_n^2}
  \bigg\{ 2 \Big(a_nN_{o}A_n^2+b_nN_{e}B_n^2+(a_nN_{e}+b_nN_o)A_nB_n\Big) \nonumber\\
  &\hspace{0pt}
  +\frac{4}{n(n+1)}\Big(c_n\bar{N}_oA_n^2+d_n\bar{N}_{e}B_n^2
  +(c_n\bar{N}_{e}+d_n\bar{N}_o)A_nB_n\Big)\bigg\}\bigg] \nonumber\\
  &\hspace{0pt}+ C_0A_0,
  \label{pw}
\end{align}
where $C_0$ is given in terms of the surface tension  of the droplet, $\gamma$,  as 
$C_0 = \{\gamma-\mu\ot A_0(2\chi^2+1)/(\chi^2-1)\}/(\pi \mu^{(1)}a^2\chi)$
based on the  condition $\Pi\ot_{rr}-\Pi\In_{rr} =2\gamma/b$.  Again, all
undefined constants are given in  Appendix A.

\section{Numerical simulations}\label{sec:numerical}
In parallel with our theoretical approach, we use numerical simulations based on a $3$D boundary element method.
By choosing  the characteristic length, velocity, and stress as $b$,  
$\lambda\gamma/\{ \mu^{(2)} (1+\lambda)\}$, and $\gamma/b$ respectively, 
the nondimensional boundary integral formulation 
 for the matching-viscosity case ($\lambda=1$) can be  obtained. 
The nondimensional velocity $\bm{u}(\bm{x}_0)$ at position $\bm{x}_0$ 
everywhere in the domain is classically written as
\begin{align}
  \bm{u}(\bm{x}_0) = \frac{1}{2\pi}\int_{\tilde{S}}\kappa(\bm{x})\bm{n}(\bm{x}) \cdot \bm{G}(\bm{x}_0,\bm{x}) dS(\bm{x})
  -\frac{1}{4\pi} \int_{\hat{S}} \bm{q}(\bm{x})\cdot \bm{G}(\bm{x}_0,\bm{x}) dS(\bm{x})
  \label{equ:bim}
,\end{align}
where 
$\tilde{S}$ and $\hat{S}$ denote the surface of the droplet and swimmer respectively,
$\bm{n}$ the normal vector on $\tilde{S}$ towards the outer fluid,
$\kappa = -\frac{1}{2}\nabla_s \cdot \bm{n}$ the mean curvature of $\tilde{S}$,
and $\bm{q}$ the density of the single-layer potential on $\hat{S}$.
The tensor $\bm{G}$ is the free-space Green's function, 
also known as the Stokeslet or the \textit{Oseen-Burgers} tensor,
\begin{align}
  \bm{G}(\bm{x}_0,\bm{x}) = \frac{\bm{\delta}}{r} + \frac{(\bm{x}_0-\bm{x})(\bm{x}_0-\bm{x})}{r^3},
\end{align}
where $\bm{\delta}$ is identity tensor and $r = |\bm{x}_0-\bm{x}|$.
As shown in Eq.~\ref{equ:bim}, only single-layer integration is performed, 
which is sufficient for the rigid body motion of the swimmer and 
the dynamics of a matching-viscosity droplet\cite{pozrikidis1992boundary}.

\begin{figure}
\centering
\includegraphics[scale=0.6]{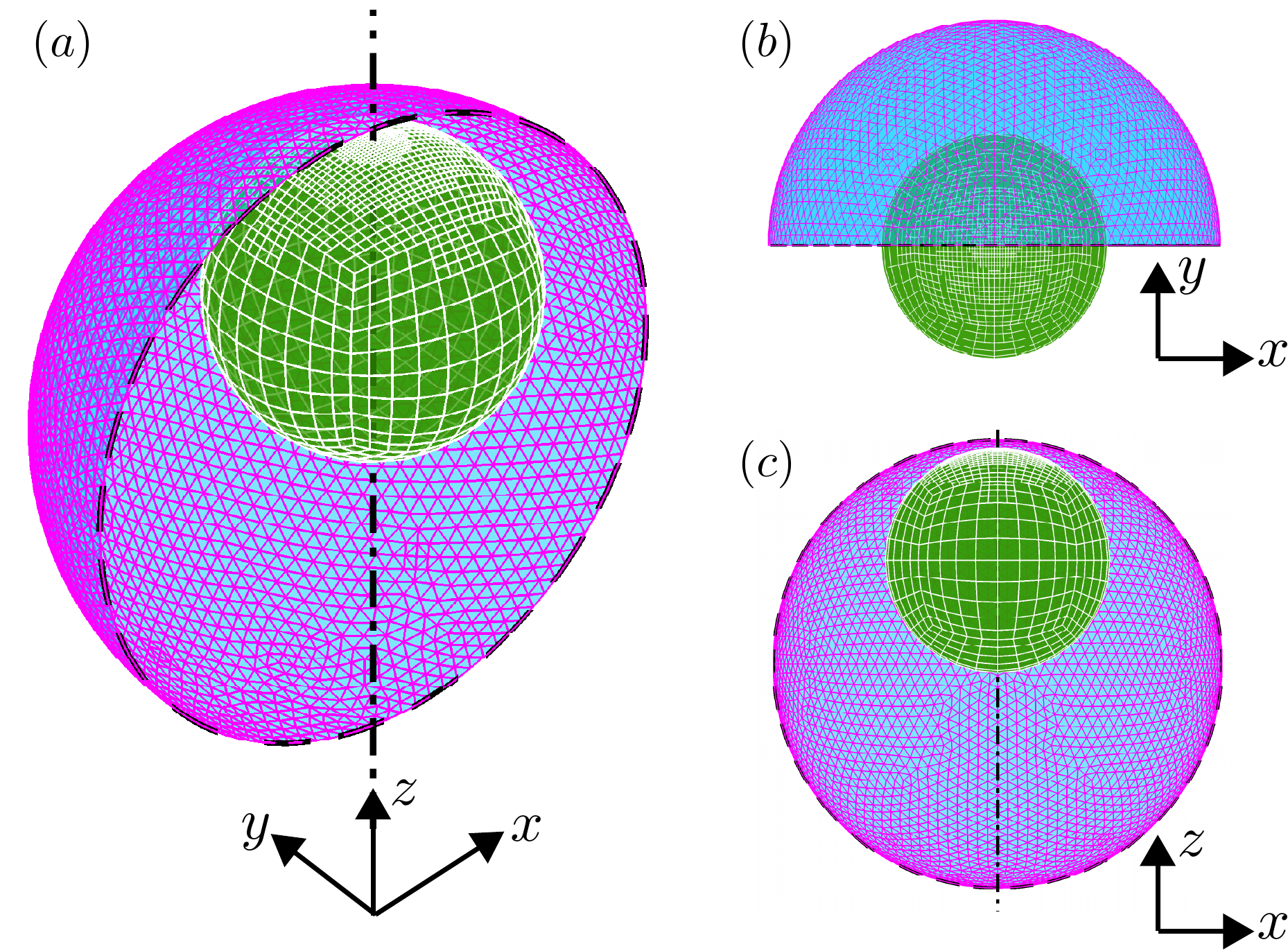}
\caption{Meshing of the swimmer-droplet pair used in numerical simulations: (a) The $3$D view of the meshes of the droplet (triangular elements) and swimmer (quadrilateral elements), 
where adaptive mesh refinement is implemented on the swimmer; 
half of the droplet interface is removed
for visualisation purposes.  (b) The projection view on the $xy$ plane.
  (c) The projection view on the $xz$ plane.}
\label{fig:mesh_bim}
\end{figure}

The surfaces of the swimmer and droplet are discretized using  
zero-order flat quadrilateral and second-order curved triangular elements respectively.
For the spherical swimmer, a six-patch structured mesh\cite{tube_higdon,zhu2013low} consisting of $600$ (before mesh 
refinement) elements is constructed.
The number of elements on the droplet interface is around 
$2500$ ($\sim 5000$ discretized points).
Gauss-Legendre quadrature is applied on the quadrilateral elements to compute nonsingular 
integrations; on triangular elements, we compute the integrations 
using a symmetric Gaussian quadrature rule\cite{dunavant1985high}. 
When $\bm{x}_0$ is on the surfaces $\tilde{S}$ or $\hat{S}$, 
the surface integrals become singular and
different desingularization strategies are chosen:
on the droplet interface $\tilde{S}$, the well-known integral identity for $\bm{G}$ is exploited and
hence the first integral in Eq.~\ref{equ:bim} becomes
\begin{align}
  \int_{\tilde{S}}\kappa(\bm{x})\bm{n}(\bm{x}) \cdot \bm{G}(\bm{x}_0,\bm{x}) dS_x
  =\int_{\tilde{S}}[\kappa(\bm{x})-\kappa(\bm{x}_0)] \bm{n}(\bm{x}) \cdot \bm{G}(\bm{x}_0,\bm{x}) dS_x,
\end{align}
where the $\mathcal{O}(r^{-1})$ singularity of the original integrand is removed;
on the squirmer surface $\tilde{S}$, each quadrilateral element is divided into four triangular sub-elements, 
where polar coordinates transformation\cite{poz_blue} with Gauss-Legendre 
quadrature is adopted to desingularize the integral. 
Both integrals in Eq.~\ref{equ:bim} tend to be nearly singular when 
the distance between the two surfaces $\tilde{S}$ and $\hat{S}$ is too small. 
Desingularizing measures are hence taken for them:
on the droplet interface $\tilde{S}$, a high-order near-singularity subtraction is implemented 
by following Ref.~\cite{zinchenko2006boundary} and
on the swimmer surface $\hat{S}$, adaptive mesh refinement is utilized.
Figure~\ref{fig:mesh_bim} presents a schematic view of the adaptively-refined mesh.


A crucial numerical difficulty arising from droplet/bubble simulations based on Lagrangian interface 
representation is to maintain the quality of the mesh of the interface. 
In order to guarantee the smoothness and orthogonality of the triangle mesh over a long time evolution, 
we implement a so-called `passive mesh stabilization' scheme
\cite{zinchenko1997novel, zinchenko2013emulsion}. 
At each time step, the scheme searches the optimal tangential field that is added to the normal velocity to update the Lagrangian points, 
minimizing a global kinetic-energy-like norm that quantifies the clustering and distortion of the mesh. 
This scheme significantly slows down mesh degradation.
Its effectiveness was proved in the previous study on a squeezed pancake droplet 
in a microfluidic chip based on an accelerated boundary integral implementation\cite{zhu16pancake}.


In contrast to the infinite-surface tension assumed in the theory, 
a  large but finite surface tension is adopted in the simulations and hence the numerical  droplet is not strictly spherical but slightly 
deformable. The strength of the typical ratio of viscous stresses to  surface tension forces is measured by the
capillary number,  $\text{Ca} \equiv \mu^{(2)} B_1 /\gamma$, and $\text{Ca}=0$ corresponds to the theoretical limit  of infinite surface tension. We vary $\Ca$ numerically from $10^{-3}$ to $10^{-2}$, without detecting significant changes in the kinematics of 
the swimmer. We hence use $\Ca=10^{-3}$ throughout our study, and are able to  approximate well the $\Ca=0$ limit from our 
\textit{a 
posteriori} comparison with the theory. 

\section{Results}

\subsection{Squirming with purely tangential velocities}~\label{sec:inst}
In this section, we start by investigating the instantaneous dynamics of a droplet encapsulating a squirmer using solely 
tangential surface velocities, {\it i.e.}~with $A_n=0$. 
If one further sets the $B_n$ ($n\ge 3$) modes to zero, as is classically done for the squirmer model \cite{ishikawa:06}, 
the swimming gait consists of only $B_1$ and $B_2$ modes:   the $B_1$ mode determines the swimming velocity while the  $B_2$ mode captures the leading order 
 disturbance flow induced by the swimmer,  namely a stresslet (or force dipole). We define $\beta\equiv B_2/B_1$   to measure the relative strength of the
stresslet. The squirmer is said to be  neutral when $\beta=0$, while it is a pusher (respectively a puller) when $\beta$ is negative (respectively positive). Varying the value of $\beta$ allows to model the majority of  swimming microorganisms and synthetic microswimmers: 
pushers model flagellated bacteria such as  \textit{E.~coli} \cite{berg04}
  while  biflagellated green algae such as 
\textit{C. reinhardtii}\cite{graham:09,spag:12,goldstein:15} are  pullers. Neutral swimmers may be 
considered as special cases of synthetic swimmers such as Janus particles self-propelling owing to various phoretic 
mechanisms or some active droplets driven by Marangoni 
stresses\cite{yoshi:12,schmitt:13,herm:14,herm:16,goles:05,anderson:89,wang:13,colberg:14}.

\subsubsection{Velocity of the swimmer and droplet: theory}\label{sec:sw}
For a tangential squirmer, the velocities $U_S$ and $U_D$ are given analytically by
\begin{align}
  &\frac{U_S}{U_0} = \frac{3}{\Delta}\{(2\chi^5-5\chi^2+3)\lambda + 3\chi^5+5\chi^2-3\}, \nonumber \\
  &\frac{U_D}{U_0}  = \frac{15\chi^2}{\Delta},
  \label{vel_bmod}
\end{align}
where $\Delta$ is defined in Eq.~\ref{eq:delta} and we use the swimming velocity  $\Uu$ of an unbounded 
squirmer as the reference scale, $\Uu=2B_1/3$. Both velocities
are functions solely 
of the size ratio, $\chi$, and the viscosity ratio, $\lambda$.

\begin{figure}[t]
  \centering
  \includegraphics[width=7.2cm,angle=0]{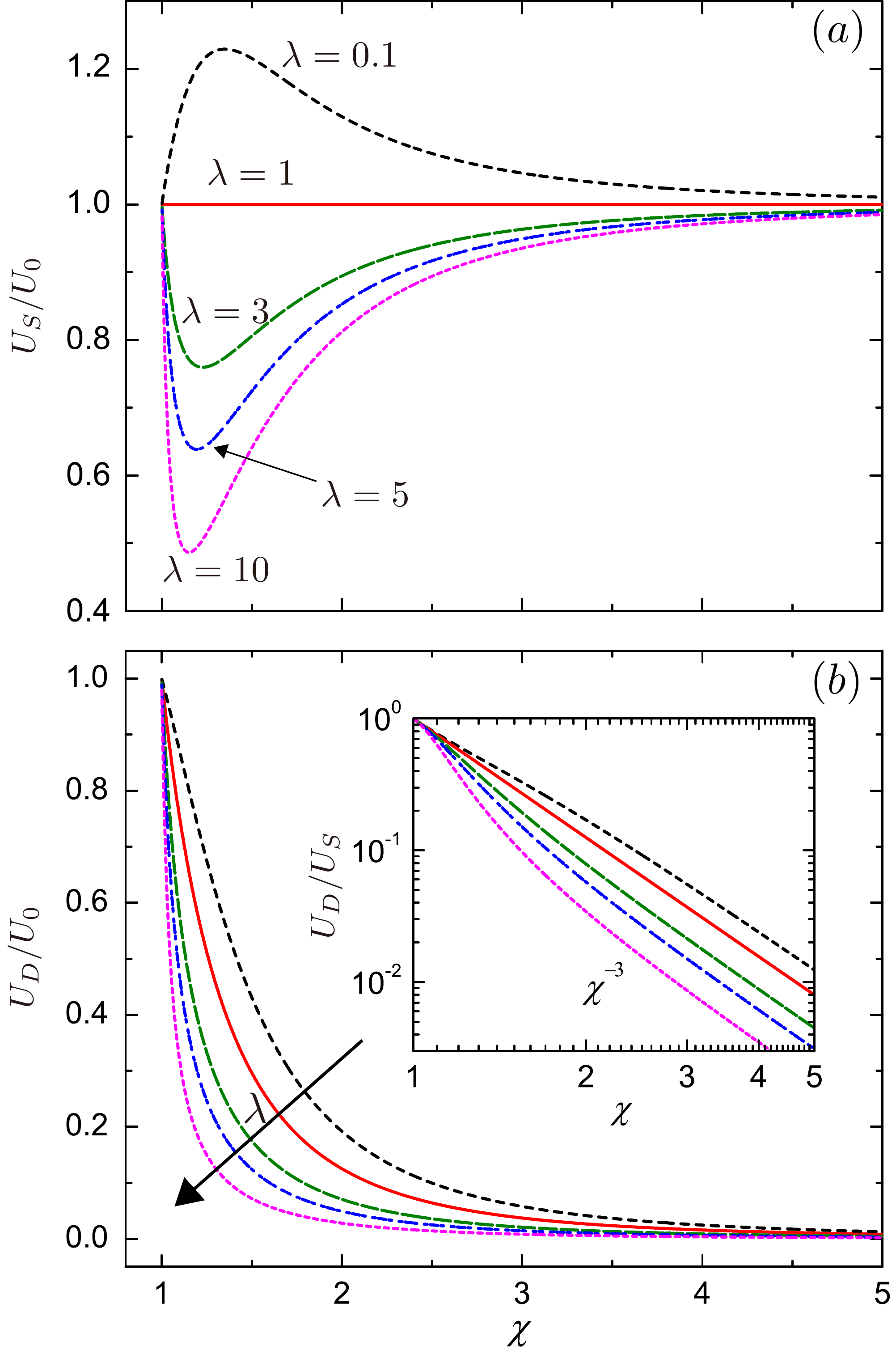}
  \caption{(a): Velocity $\Us$  of the swimmer squirming with tangential surface actuation only;  (b): Velocity $\Ud$ of the 
droplet, scaled by that of an unbounded squirmer, $\Uu=2B_1/3$. In both cases, the velocities are plotted  
  as a function of the size ratio, $\chi=b/a$ for
  viscosity ratios $\lambda=0.1$, $1$, $3$, $5$ and $10$.
 The inset of (b) shows the ratio, $\Ud/\Us$, of the droplet velocity over the swimmer velocity
 in log-log form. 
  }
  \label{vel_lam}
\end{figure}

We plot in Fig.~\ref{vel_lam}a the dependence of the swimmer velocity $\Us$ on $\chi$ 
and $\lambda$. The velocity  decreases monotonically
with $\lambda$. When the outer and inner phase have  matching viscosities ($\lambda=1$),
$\Us$ is not affected by the presence of the droplet, and is  thus equal to the unbounded velocity $\Uu$ for all values of $\chi$. The 
squirmer 
swims faster than the unbounded one when the outer phase is less viscous than the inner ($\lambda<1$), and  swims 
slower in the opposite limit, $\lambda>1$. When $\lambda \neq 1$, the velocity $\Us$ varies with the size ratio $\chi$ non-monotonically, reaching its maximum value for $\lambda<1$ when the swimmer is  tightly confined, $\chi \approx 1.1 \sim 1.2$, namely when the droplet is slightly larger than the swimmer. The result is similar when  $\lambda>1$ and the minimum is reached.  For any viscosity ratios, $\Us=\Uu$ in the limit of $\chi=1$ and $\chi\rightarrow\infty$. The former corresponds to 
the situation when the droplet exactly encompasses the swimmer and the latter to  when the droplet is much 
larger than the swimmer. In Fig.~\ref{vel_lam}b, we further show that the velocity $\Ud$ of the droplet decreases 
monotonically
with $\chi$, as well as  with $\lambda$. The inset of Fig.~\ref{vel_lam}b presents the 
ratio $\Ud/\Us$ of the droplet velocity over the swimmer velocity as a function of $\chi$ in a log-log form, this ratio 
decays as $\chi^{-3}$ for large $\chi$. It is important to note that for any values of $\chi$ or $\lambda$ the swimmer is always 
faster 
than the droplet, $\Us>\Ud$. The concentric configuration is thus not a steady state if the swimmer only applies tangential forcing.

\subsubsection{Comparisons between theory and 
simulations}\label{sec:comp}

\begin{figure}[t]
  \centering
  \hspace{-0.83cm}\includegraphics[width=8.28cm,angle=0]{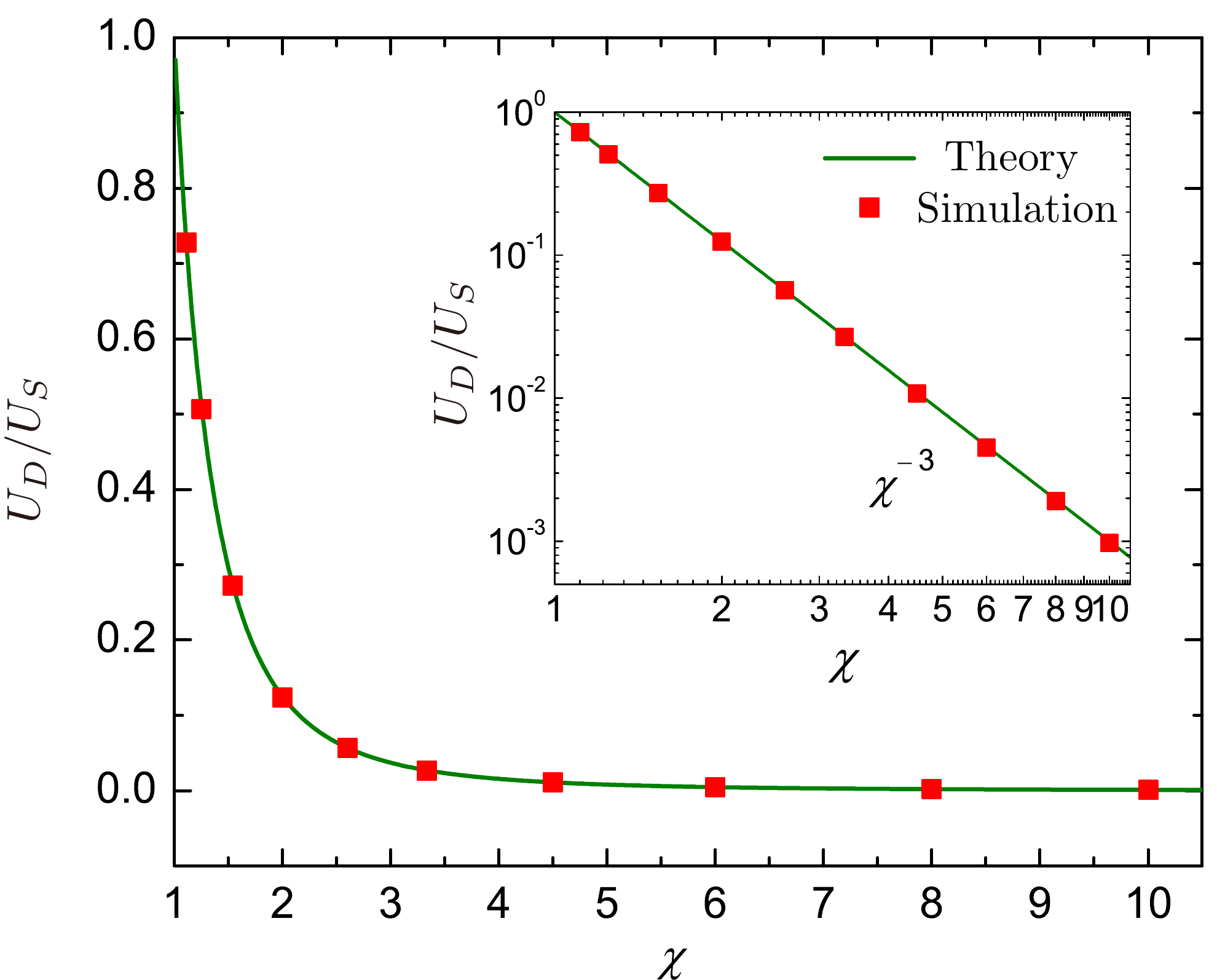}
  \caption{The ratio $\Ud/\Us$ between the droplet velocity, $\Ud$, and the swimmer velocity, $\Us$, as a function of  the size 
ratio $\chi$. The swimmer employs only tangential squirming modes  and the viscosity ratio is $\lambda=1$.  Green solid lines   and red  squares indicate  results from the theory and numerical simulations,  respectively.
  }
  \label{vel_ratio}
\end{figure}

Here we consider the dynamics of a neutral swimmer ($\beta=0$), a  pusher with $\beta=-5$ and a puller 
with $\beta=5$ encapsulated 
inside a same-viscosity droplet ($\lambda=1$). For simplicity we further take $B_n=0$ for $n\geq 3$ and $A_n=0$ for all $n$. 
Since the velocities $\Ud$ and $\Us$ are independent of $\beta$, 
the ratio $\Ud/\Us$ only depends on the value of $\chi$. This  
functional dependence is plotted in  
 Fig.~\ref{vel_ratio}, showing an excellent  agreement between the theory (green lines) and numerical data (red squares).

\begin{figure*}
  \centering
  \includegraphics[scale=0.5,angle=0]{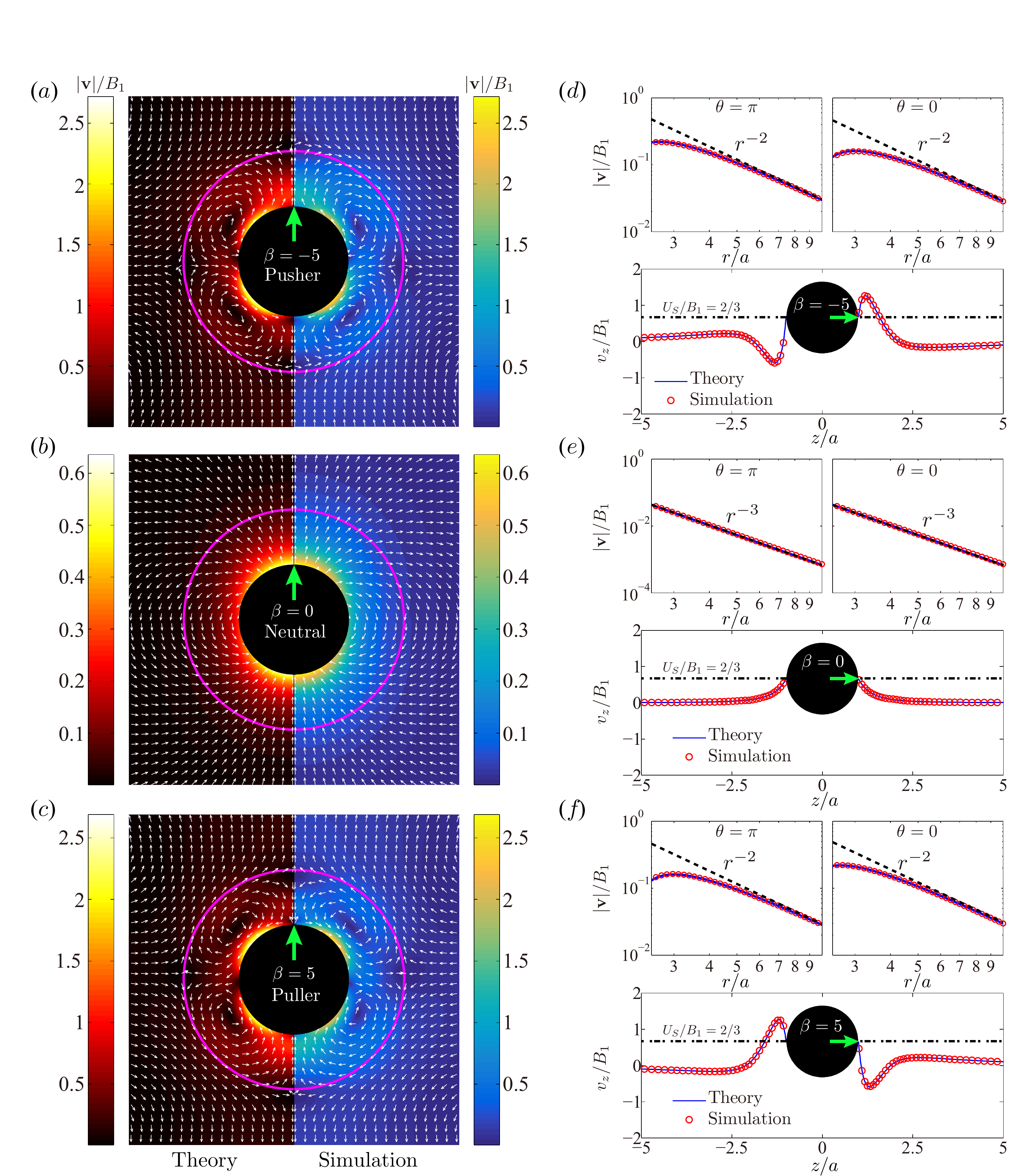}
  \caption{Illustration of velocity fields, $\bm{v}$, in the laboratory frame.
  Comparison between the theory and simulations
    for a pusher with $\beta=-5$ (a, d), a neutral swimmer with $\beta=0$ (b, e), 
    and a puller $\beta=5$ (c, f). The viscosity ratio is $\lambda=1$ and the 
size ratio is 
$\chi=2$. Black spheres 
denote the swimmers and solid magenta circular lines the droplets. The  green 
arrows 
indicate the swimming directions.
    The left column (a, b, c) display the velocity vectors (white arrows) of
    $\bm{v}/B_1$ and the contours of its 
magnitude $|\bm{v}|/B_1$.  Theoretical  results are  shown on
    the left panels while the numerical results are shown on the  right. 
    The right column (d, e, f) shows the theoretical (blue solid line) and 
    numerical (empty red circles) data of the scaled velocity along the $z$ 
axis, 
$v_z/B_1$,  where the dot-dashed line indicates the swimmer velocity of $\Us/B_1=2/3$. 
The velocity magnitude, $|\bm{v}|/B_1$, versus the distance, $r/a$, along
the anterior $\theta=0$ ($z>0$) and 
posterior 
$\theta=\pi$ ($z<0$) directions is shown in a log-log form; the dashed curves
denote the leading order 
velocity $v_r\ot|_{\mathrm{leading}}$ at $\theta=0$ and $\pi$.
The spatial decay of $|\bm{v}|/B_1$ in the far field follows the $r^{-2}$ law 
for the pusher/puller and $r^{-3}$ law 
for the neutral swimmer.
  }
  \label{vel_field}
\end{figure*}

Next in Fig.~\ref{vel_field}a-c, we  plot the flow velocity field, $\bv/B_1$, 
in the laboratory frame for the pusher (a), 
neutral (b)  and puller (c) swimmers  respectively. The size ratio is $\chi=2$.
Theoretical results are shown on the left panel and numerical data on the right.  The numerical predictions show  good agreement with the theoretical data in most of the flow domain except very close to the  droplet interface where numerical errors arise from the nearly-singular integration.

For the neutral swimmer,  note that the velocity field 
is not affected at all by the presence of the droplet.  This is corroborated by the fact that neither the swimming velocity nor the power are 
impacted by the droplet, as implied by Eq.~\ref{vel_sw} and Eq.~\ref{pw}. 
This results from the vanishing radial velocity 
in the droplet frame, such that the spherical droplet interface 
introduces no perturbation and hence does not influence the swimming dynamics.

For the pusher in a drop,  similarly to  a pusher in free space, fluid is locally pushed away from the anterior ($\theta=0$) 
  and posterior ($\theta = \pi$) parts of the swimmer and comes to 
the lateral directions ($\theta = \pi/2$).
Due to the non-penetrating nature of the droplet interface, 
two counter-rotating toroidal vortices form inside.
Outside the droplet the fluid is drawn towards its
poles and expelled away on the equatorial plane. 
Interestingly the flow signature of a local pusher turns therefore into a puller in a far field. More quantitatively one can show that  the velocity fields of a puller with 
$\beta >0$ and a pusher with $-\beta$
satisfy the relation
\begin{align}\label{eq:img}
v_r|_{\beta} \lp r,\pi-\theta \rp + v_r|_{-\beta}\lp r,\theta \rp = 0, \nonumber\\
v_{\theta}|_{\beta} \lp r,\pi-\theta \rp - v_{\theta}|_{-\beta}\lp r,\theta \rp = 0,
\end{align}
which indicates that the mirror symmetry about the equatorial plane
$\theta=\pi/2$ of the flow field of the pusher 
with $-\beta$
is equivalent to the reversed flow field of the puller with $\beta$.

We next investigate the spatial variation of 
$\bv \lp z \rp /B_1$ along the $z$ axis  in Fig.~\ref{vel_field}d, e and f for 
the three swimmers. Here again,  
numerical data (empty red circles) agree very well with the theoretical 
predictions (solid bule line). The velocity magnitude, 
$|\bv|/B_1$, decays in the far-field from the swimmer 
center as $r^{-2}$ for the pusher/puller and $r^{-3}$ for 
the neutral swimmer. The velocity distribution $\bv \lp z \rp$ over $z$ for the 
pusher and that for the puller are symmetric about $z=0$, as implied by 
Eq.~\ref{eq:img}. For both swimmers, two stagnation points appear near the 
droplet interface $r=b$, one close to the frontal interface and other close to the rear. They can be observed 
 in Fig.~\ref{vel_field}.

It is worth emphasizing the result   that the presence of droplet reverses 
the direction of the far-field flow with respect to that of a pusher/puller in 
free space (Fig.~\ref{vel_field}a and c). This can be made more precise by an 
analysis of the theoretical predictions in  Eq.~\ref{vel_full}. 
With only $B_1$ and $B_2$ modes, the leading-order contribution to the radial
velocity $v_r\ot$ in the outer phase is 
\begin{align}
 v_r\ot|_{\mathrm{leading}} = -\frac{c_2}{\Delta_2 \chi^2}\lp\frac{a}{r}\rp^2 B_2 P_2\lp\xi \rp,
\end{align}
and that to the radial velocity of an unbounded pusher/puller is given by Ref.~\cite{blake:71} as
\begin{align}
 v_r|_{\mathrm{leading}} = - \lp\frac{a}{r}\rp^2 B_2 P_2\lp \xi \rp .
\end{align}
Their ratio is $v_r\ot|_{\mathrm{leading}}/v_r|_{\mathrm{leading}}=c_2/\lp\Delta_2 \chi^2\rp$, which 
is negative for any size ratio $\chi>1$ hence rationalizing  the velocity inversion.

\subsubsection{Power consumption}\label{sec:comp}

When the viscosities inside and outside the droplet are equal ($\lambda=1$) 
and the swimmer uses tangential surface actuations alone, 
the power consumption $\mp$ based on Eq.~\ref{pw} is simplified to
\begin{align}
  \frac{\mp}{4\pi {\mu}_1 a} = \frac{4}{3}B_1^2 
  + \sum_{n\ge 2}^{\infty}\frac{8\bar{d}_n}{n(n+1)\bar{\Delta}_n}B_n^2,
  \label{pw_eqvisc}
\end{align}
where
\begin{align}
  &\bar{d}_n = 4\chi^{2n+3}-(2n+3)\chi^4+(2n-1),\nonumber\\
  &\bar{\Delta}_n=8\chi^{2n+3}-(2n+1)(2n+3)\chi^4 \nonumber\\
  &\hspace{25pt}+ 2(2n-1)(2n+3)\chi^2 -(2n-1)(2n+1).
\end{align}

Restricting then our attention to the simplest squirmer with $B_n=0$ for $n \ge 3$, 
the power becomes
\begin{align}
  \frac{\mp}{4\pi {\mu}_1 a B_1^2} = \frac{4}{3} \left(1 + 
  \frac{4\chi^{7}-7\chi^4+3}
  {8\chi^{7}-35\chi^4 + 42\chi^2 -15} \beta^2\right),
  \label{pw_th1}
\end{align}
of a similar form to that of an unbounded squirmer\cite{blake:71} 
\begin{equation}
\frac{\mpu}{4\pi {\mu}_1 a B_1^2}=\frac{4}{3} \lp 1 +\frac{1}{2} \beta^2 \rp.
\end{equation}

\begin{figure}[t]
  \centering
  \includegraphics[width=7.35cm,angle=0]{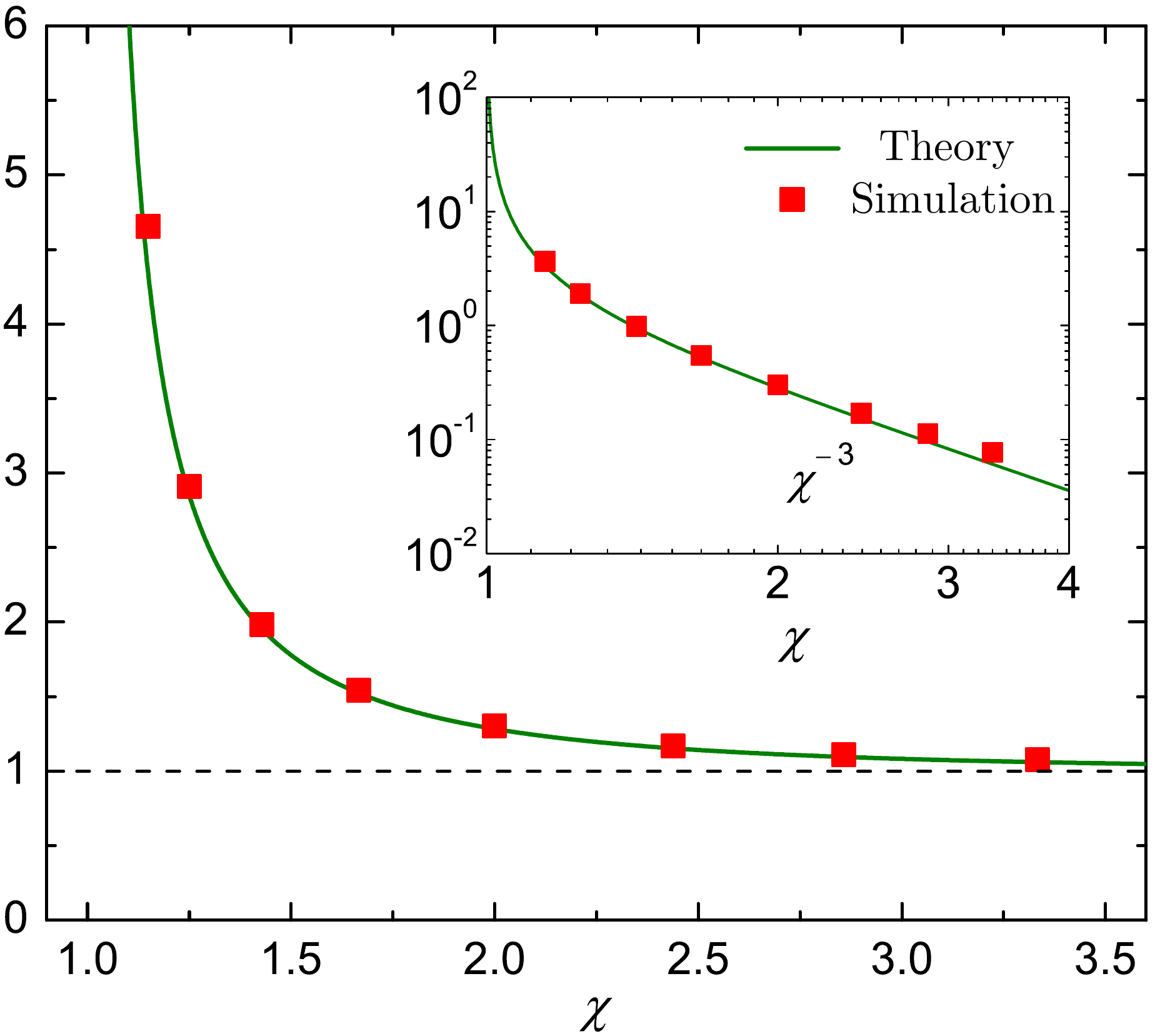}
  \put(-230,85){\rotatebox{90}{$\mp/\mpu$}}
    \put(-147,108){\rotatebox{90}{$\mp/\mpu-1$}}
  \caption{Similar to Fig.~\ref{vel_ratio}, but for the power 
  consumption  of the squirmer, $\mp$,  scaled by the unbounded 
value, $\mpu$. In contrast to the velocities, $\mp$ depends also
on modes $|B_n|$ ($n \ge 2$). Here $|\beta|=|B_2/B_1|=5$
and $B_n=0$ ($n \ge 3$). The inset shows the $\chi^{-3}$ scaling of the 
nondimensional excess power 
$\mp/\mpu - 1$. }
  \label{pow}
\end{figure}

Theoretical and numerical values of $\mp$ show excellent agreement, 
as shown in Fig.~\ref{pow}. The power  of 
an encapsulated squirmer, $\mp$,  always 
exceeds that of an unbounded one, $\mpu$. From a practical standpoint,  
$\mp$ approximately doubles when the radius of the droplet is $50\%$ 
larger 
than that of the swimmer. We further observe that
$\mp$ is negatively correlated  to $\chi$, and  the swimmer expends more
energy due to a stronger confinement. The 
inset log-log plot indicates that scaled excessive power $\mp/\mpu-1$
decreases with the size ratio  as $\chi^{-3}$.

\subsection{Co-swimming by combining   tangential and normal
  squirming}\label{sec:comov}

We have shown in the previous sections that
a swimmer employing solely tangential squirming modes, $B_n$, is always
faster than the droplet, \textit{i.e.} $\Us>\Ud$. Thus, the swimmer and 
droplet cannot remain  concentric. With the idea of using  artificial swimmers
encapsulated in a droplet for controllable cargo delivery, it is 
attempting to try and tune the squirming gait such that the swimmer and droplet co-move 
with a same velocity $\Us = \Ud$ and maintain a concentric configuration.  
We find that a squirmer combining  both tangential and 
normal velocities is able to accomplish this, as shown below.

The results in Eq.~\ref{vel_sw} and \ref{vel_drop} imply that the swimming velocities
$\Us$ and $\Ud$ only depend on the first modes, $A_1$ and $B_1$. We define
$\alpha \equiv A_1/B_1$ to indicate the relative strength of the  modes. 
By comparing Eq.~\ref{vel_sw} and \ref{vel_drop}, we find that a
particular value of $\alpha$, denoted by $\alco$,  allows to obtain equal 
velocities, namely
\begin{align}\label{al_co}
  \alco = \frac{\lp 4\lambda+6\rp \chi^5 - 10\lambda \chi^2 + 6\lp \lambda-1 \rp }{\lp 
2\lambda+3\rp \chi^5 + 10\lambda \chi^2 - 12\lp \lambda-1 \rp},
\end{align}
leading to a co-swimming squirmer and droplet velocity, $\Uco$, given by
\begin{align}
 \Ud = \Us = \Uco = \frac{10B_1 \chi^2 \{ \lp 6\lambda + 9 \rp \chi^5 -6\lp \lambda-1 \rp \}}{\Delta \{ \lp 2\lambda + 
3 \rp \chi^5 +10\lambda \chi^2-12\lp \lambda-1 \rp \}}\cdot
\end{align}
For any size ratio $\chi>1$, $\alco>0$ and thus a positive 
$A_1$ mode, which contributes to the swimming velocity negatively and 
therefore enables the squirmer to co-swim with the droplet.

\begin{figure}[t]
  \centering
  \hspace{0cm}\includegraphics[width=10cm,angle=0]{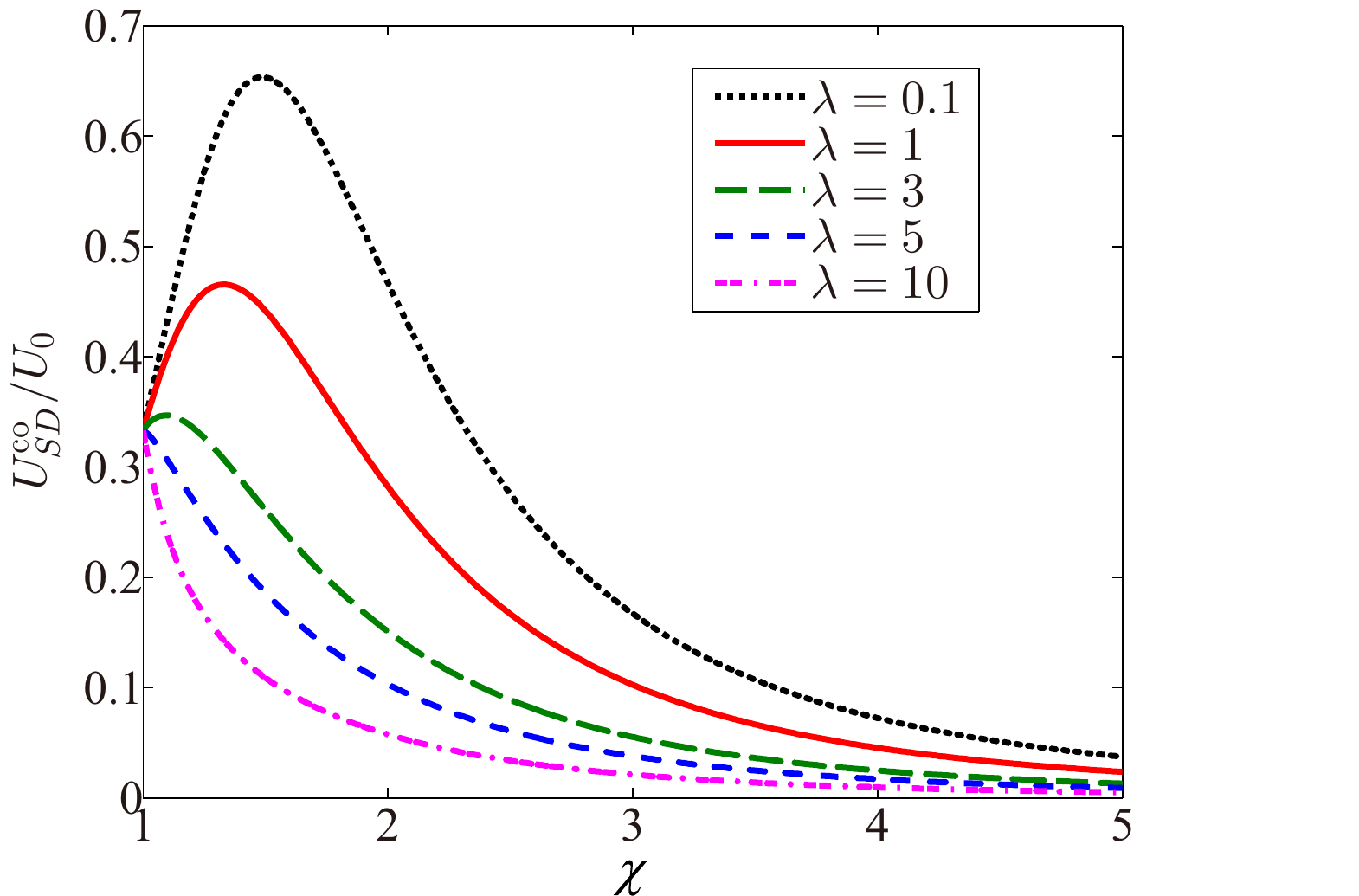}
  \caption{The co-swimming velocity $\Uco$ of the squirmer and droplet, as a function
  of the size ratio $\chi$ and viscosity ratio $\lambda$. The first-mode normal squirming
  is tuned to be $A_1 = \alco B_1$ such that the squirmer and droplet swim with a same velocity $\Uco$ .
  }
  \label{Uco}
\end{figure}

The influence of confinement $\chi$ and viscosity ratio $\lambda$ on the resulting co-swimming speed is depicted in Fig.~\ref{Uco} by plotting the scaled co-moving speed $\Uco/\Uu$,
where $\Uu = 2B_1/3$ is the velocity of an unbounded 
squirmer with pure tangential modes. Even for small viscosity ratio ($\lambda=0.1$),
the co-moving velocity $\Uco$ of the pair remains below $0.7\Uu$. Simulations 
have
been performed to determine the values of $\alco$ and 
$\Uco$ for the $\lambda=1$ case, and here again the numerical results show
excellent agreement with the theory (not shown).

The relation between the mode strength $\alpha$
and the size ratio $\chi$ required to achieve concentric co-swimming is given by Eq.~\ref{al_co} for arbitrary viscosity ratio $\lambda$. 
 When $\lambda$ is fixed,  the particular value $\alco$ ensuring co-swimming is easily chosen as a function of $\chi$. 
Conversely, one may determine a particular size ratio $\chi^{\mathrm{co}}$ as a function
of $\alpha$  by solving the quintic equation. 
In the case of $\lambda=1$, the required size ratio $\chi^{\mathrm{co}}$ is simply given by
\begin{align}\label{dropletchoose}
  \chi^{\mathrm{co}}= \lp \frac{\alpha+1}{\alpha-1/2} \rp^{1/3}.
\end{align}
It implies that for a given swimmer with fixed modes 
one may select a particular size of droplet  
transportable by the swimmer in a co-swimming state. This encouraging result  
points to a practical route toward building self-propelled chemical droplets.

\subsection{Stability of co-swimming state: axisymmetric configuration}\label{sec:stability}
\begin{figure*}
\centering
\includegraphics[scale=0.3]{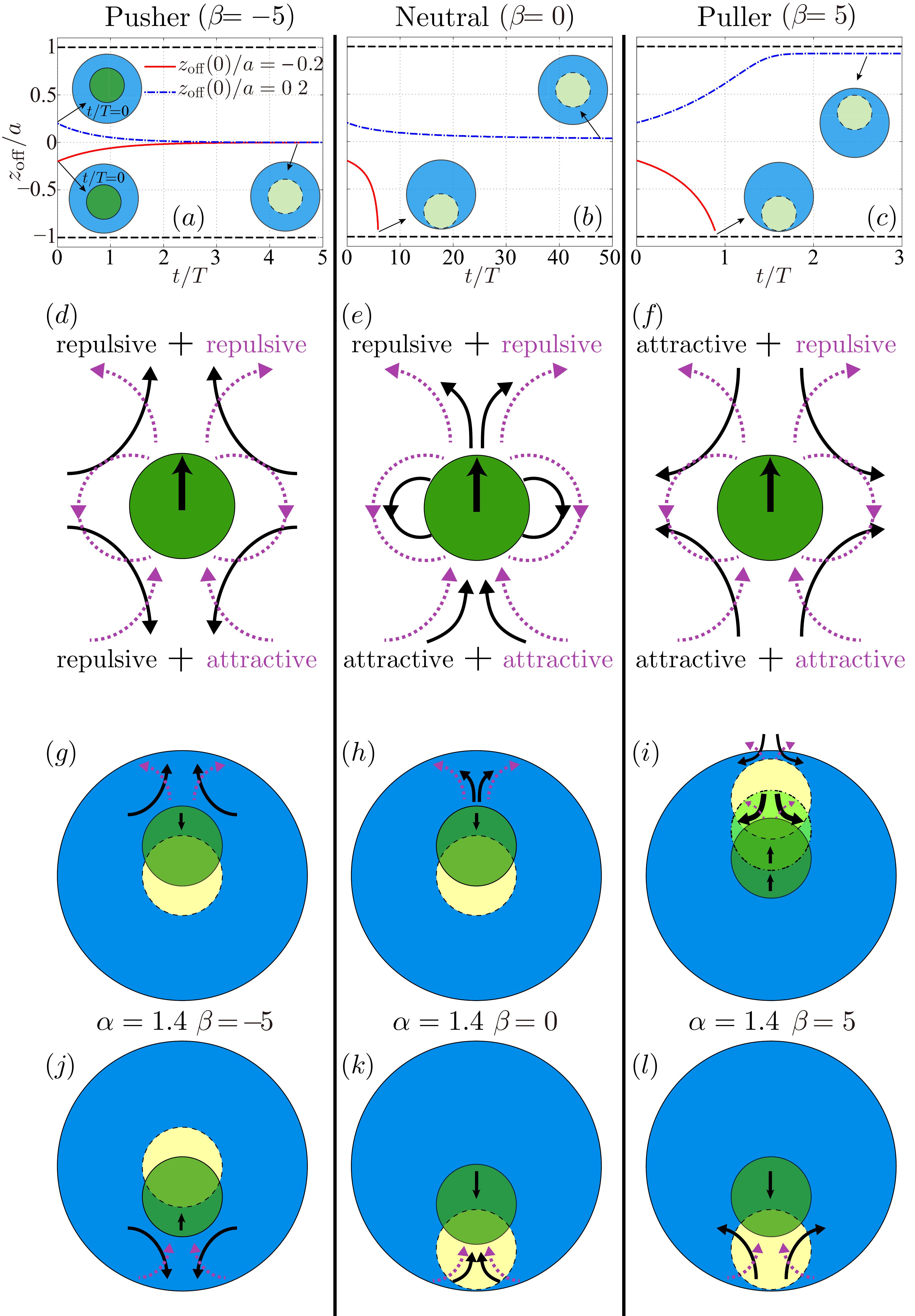}
\caption{Stability of co-swimming state. Top:
Time evolution of the axial offset position  $z_{\textrm{off}}$ of a swimmer with a co-moving swimming gait
$\alco=1.4$ with added  tangential squirming with
  (a): $\beta=-5$ (pusher); (b): $\beta=0$ (neutral); and (c): $\beta=5$ (puller).
  The swimmer is ahead/behind of the droplet center by $0.2a$ at $t/T=0$ 
  in the top ($z_{\textrm{off}}>0$)/bottom ($z_{\textrm{off}}<0$) row. 
  The  horizontal lines $z_{\textrm{off}}/a=1$ and $-1$ indicate where  the swimmer  touches the front and rear of the droplet interface respectively. 
  The solid and dashed circles indicate the swimmer's initial and final positions respectively.
Middle: Disturbance flow field induced by a swimmer with a co-moving swimming gait
    that superimposes a normal squirming of $\alco=1.4$ onto a tangential squirming of 
    (d): $\beta=-5$ (pusher); (e): $\beta=0$ (neutral); and (f): $\beta=-5$ (puller).
    The solid black and dashed magenta lines denote the flow patterns generated by the tangential and normal squirming gaits
    respectively.
Bottom:  Influence of the disturbance flows and resulting hydrodynamic interactions on the behavior of 
  a co-moving  pusher (g, j),  neutral (h, k), and  puller swimmer (i, l). 
  The green solid  sphere indicates the initial location of the swimmer while the yellow dashed circle its final location (the green dot-dashed circle in \ref{fig:flow_stab}i indicates an intermediate location.)
}
\label{fig:stability}
\label{fig:gait}
\label{fig:flow_stab}
\end{figure*}

While the analysis above shows that co-swimming is possible, 
it is not clear {a priori} if such configuration would be stable. 
In order to address the stability of swimmers, we perform  numerical simulations  for a swimmer-droplet pair   which are  initially off-center
but axisymmetric.
The stability problem depends on many parameters including the size ratio $\chi$, the
viscosity ratio $\lambda$, the value of the mode ratio $\alco$, 
the stresslet strength $\beta$, and the initial offset distance $z_{\textrm{off}}$. 
In order to make the problem tractable, we restrict the parameter values
as  $\chi=0.5$, $\lambda=1$, $\alco=1.4$  and $\beta=-5, 0, 5$. 
We use $z_{\textrm{off}}=z_{\textrm{sq}}-z_{\textrm{dp}}$ 
to denote the offset distance in the axial direction, where $z_{\textrm{sq}}$ and 
$z_{\textrm{dp}}$ are the axial positions of the swimmer and 
droplet respectively and all simulations start with $z_{\textrm{off}}(t=0)=\pm 0.2a$. 

Figure~\ref{fig:stability} (top row) displays the time evolution of $z_{\textrm{off}}$ for a swimmer which starts initially ahead (blue dot-dashed lines) 
or behind (red solid lines) using a tangential squirming of 
$\beta=-5$ (pusher, a), 
$\beta=0$ (neutral, b) and $\beta=5$ (puller, c).  The 
physical characteristic time $T=b/B_1$ is used to scale the time $t$.
For the co-moving pusher as shown in Fig.~\ref{fig:stability}a, the offset $z_{\textrm{off}}(0)$ decays to zero 
regardless of its sign: the concentric co-moving state is recovered and 
remains stable. The influence of $z_{\textrm{off}}(0)$ for the 
co-moving neutral swimmer is shown in Fig.~\ref{fig:stability}b. 
The concentric co-movement is seen to be stable if the swimmer is initially ahead of the droplet, 
but it is unstable and yields a finite-time collision between 
the swimmer and the droplet interface, when the swimmer is initially behind. 
In contrast, for the puller  illustrated in Fig.~\ref{fig:stability}c, 
the swimmer eventually touches the rear interface 
indicating instability when $z_{\textrm{off}}(0)<0$, 
while when $z_{\textrm{off}}(0)>0$, the pair reaches an eccentric
co-moving state that is 
asymptotically stable. In the later case, the swimmer is close to 
the front droplet interface but separated by a thin 
lubrication film  which acts to stabilize their co-movement via hydrodynamic interactions. 
The asymptotically steady 
thickness of the film is about $0.08a$.

The stability properties of the co-moving state seen in Fig.~\ref{fig:stability}
may be interpreted physically by examining the disturbance flow field 
induced by the swimmer. We plot in Fig.~\ref{fig:gait} (middle row)  
the disturbance flow patterns corresponding to
the co-moving swimming gaits which consist of normal squirming $\alco$ 
(dashed magenta lines) and tangential squirming $\beta$ (solid black lines).
The disturbance flow of the pusher and puller is characterized by a stresslet 
oriented in the swimming direction, decaying as $1/r^2$; that of the neutral swimmer resembles a 
source dipole along the same direction, decaying faster as $1/r^3$. The analysis of 
Ref.~\cite{blake:71} shows that the flow induced by the $A_1$ mode squirming is 
equivalent to that by a neutral swimmer with $B_1 = A_1$. 
The details of this disturbance flow dictate  hydrodynamic interactions  between 
the swimmer and its environment. 
As can be seen in Fig.~\ref{fig:gait}d, a body located in front of or 
behind a pusher tends to be repelled by it while it will tend to be attracted 
for a puller. 
In contrast for a neutral swimmer with  $A_1>0$, ahead of the swimmer will be 
repulsive while it will tend to be attractive behind it. 

We then link in Fig.~\ref{fig:flow_stab} (bottom row)  the disturbance flow
of the swimmer and its relative movement with respect to 
the droplet, where solid/dashed circles denote the swimmer's initial/final 
location (the dot-dashed circles denotes an
intermediate position). As seen  in Fig.~\ref{fig:flow_stab}g, for a co-moving
pusher  initially ahead of the droplet 
center, the repulsive flow in front of the swimmer, consisting of both repulsive
flows from tangential squirming of $\beta=-5$ and normal squirming of 
$\alpha=1.4$, 
is  stronger than its rear counterpart and brings the swimmer back to the center
(stable).  For the same swimmer but initially closer to the rear of the droplet 
 as depicted in 
Fig.~\ref{fig:flow_stab}j, the rear flows dominate. While the flows   induced 
by 
the two squirming modes are of opposite sign, the repulsive flow arising from  
tangential squirming is likely to overcome the attractive one of the normal 
squirming due 
to the faster-decaying and shorter-ranged disturbance flow  of the latter ($1/r^3$ vs.~$1/r^2$). 

The behavior of the co-moving neutral swimmer  can be   
understood along the same vein, as illustrated in Fig.~\ref{fig:flow_stab}h and 
k,
and similarly for the  puller when its is initially located behind that of the 
droplet   (Fig.~\ref{fig:flow_stab}l). The only non-intuitive result is the 
asymptotically-stable eccentric location of  
the co-moving puller that is originally closer to the droplet front as 
illustrated
in Fig.~\ref{fig:flow_stab}i. Initially, the gap between the swimmer and 
interface is relatively large, therefore the longer-ranged attractive flow  
from the tangential squirming  will outweigh the shorter-ranged repulsive one 
from the normal squirming, and the swimmer 
will be
attracted towards the interface. As the gap width decreases, the repulsive 
short-range flow 
becomes stronger, eventually dominating and preventing the  swimmer from 
further approaching the interface. This explains, at least qualitatively, why 
hydrodynamic interactions lead in this situation to  a stable eccentric 
configuration.

\begin{figure}
\centering
\includegraphics[scale=0.6]{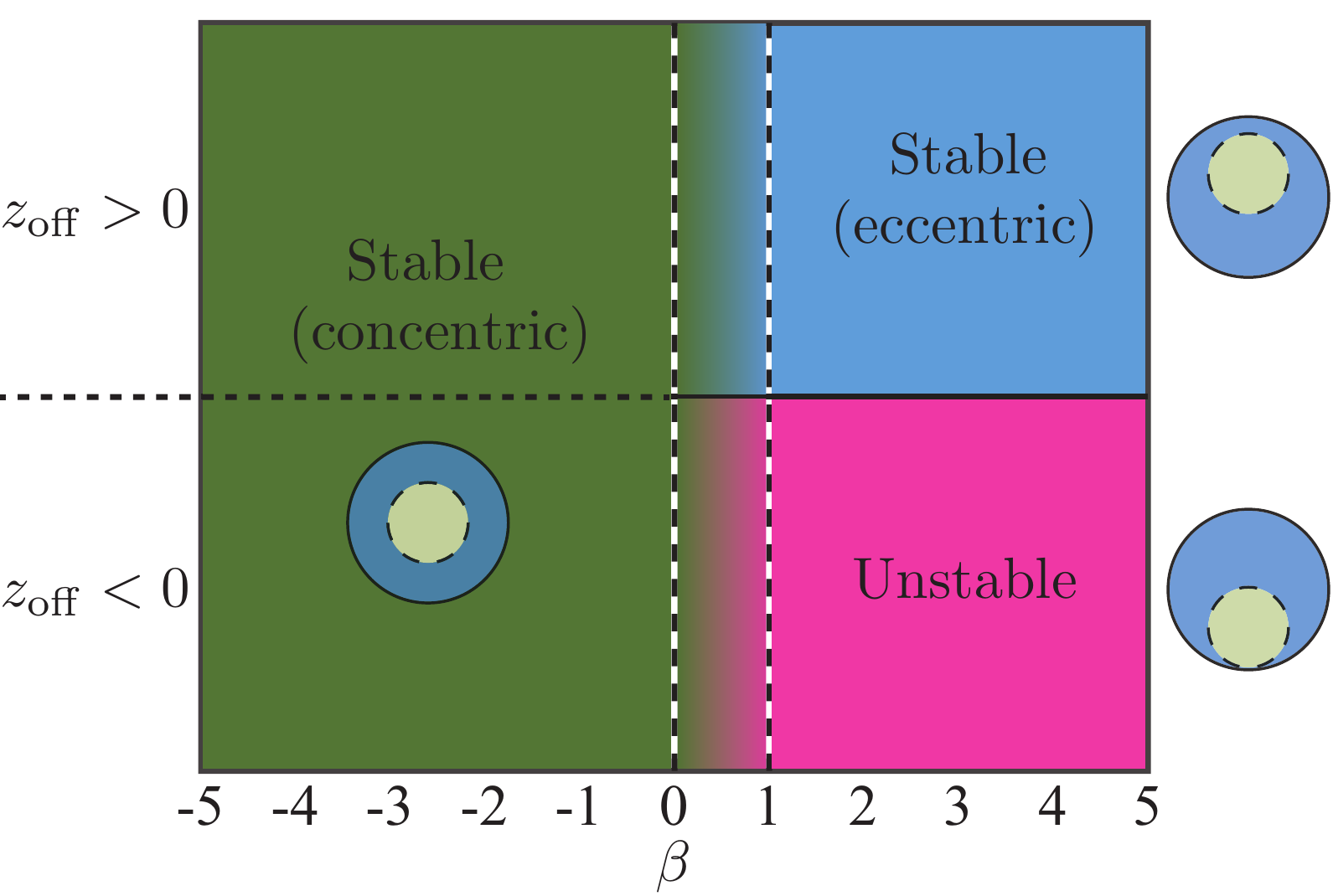}
\caption{The dependence of the stability of the co-moving state on the 
stresslet strength $\beta$. }
\label{fig:phase}
\end{figure}

Additional simulations were then performed with $1/\chi$ ranging from 0.3 to 0.7 and    $\beta$ ranging from $-5$ to $5$. These simulations show that the stability properties of the 
co-moving state is 
independent of the size ratio $\chi$ and depend only on $\beta$. As shown in 
Fig.~\ref{fig:phase}, when $\beta \leq 0$, the concentric co-movement state is 
stable regardless of the sign of the initial offset $z_{\textrm{off}}$. When 
$\beta \geq 1$, the eccentric co-moving state is stable if the swimmer 
is initially ahead ($z_{\textrm{off}}>0$) while no stable co-moving configuration 
is observed otherwise ($z_{\textrm{off}}<0$).

\subsection{Stability of co-swimming state: non-axisymmetric configuration\label{sec:stability2}}
We next address the issue of stability when the initial position of the swimmer center is not aligned with 
the droplet  along the $z$ axis.
Since the system is not axisymmetric in this case, we employ
numerical simulations allowing the swimmer to display rotational motion.
We track the  two offset distance in $x$ and $z$ directions
with $x_{\textrm{off}}=x_{\textrm{sq}}-x_{\textrm{dp}}$ and
$z_{\textrm{off}}=z_{\textrm{sq}}-z_{\textrm{dp}}$.
When $\chi=2$ and $\alco=1.4$, we consider three types of swimmers, namely
a pusher with stresslet strength $\beta=-5$, a neutral swimmer with $\beta=0$,
and a puller with $\beta=5$.

\begin{figure}[!ht]
\centering
\includegraphics[scale=0.6]{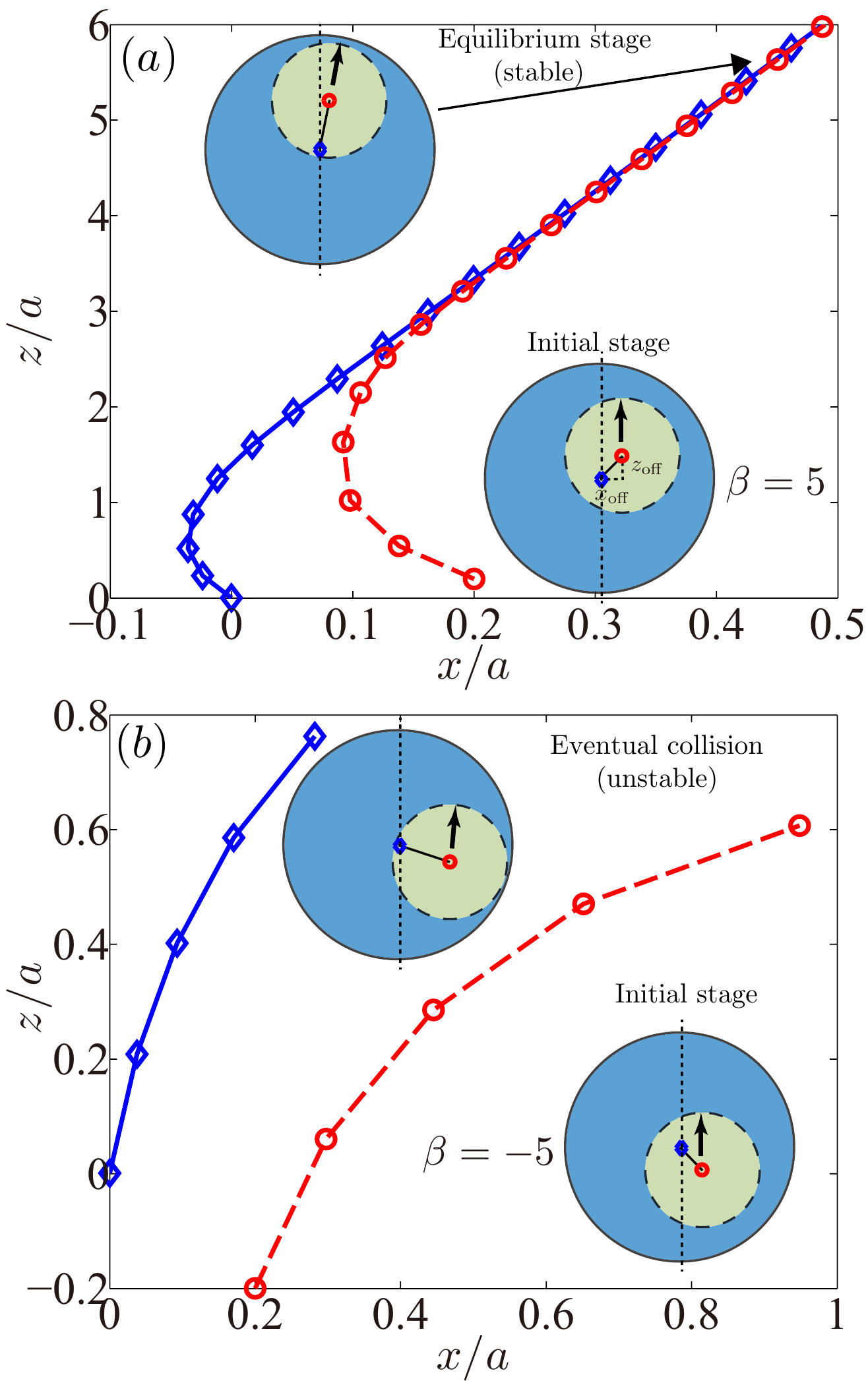}
\caption{Trajectories of swimmers in droplets initially in non-axisymmetric configurations shown in the laboratory frame: (a) pullers ($\beta=5$) with an initial offset     $(x_{\textrm{off}},z_{\textrm{off}})=(0.2a,0.2a)$
    and (b) pushers ($\beta=-5$) with an initial offset
    $(x_{\textrm{off}},z_{\textrm{off}})=(0.2a,-0.2a)$ .
    The blue diamonds and red circles denote the droplet and swimmer centers respectively.
    The arrows indicate the swimming directions.
    The puller with the initial configurations in (a) has a stable configuration
    while other swimmers collide with the droplet surface.
}
\label{fig:tra_off}
\end{figure}

We first plot in Fig.~\ref{fig:tra_off}a  the trajectories of pullers in the laboratory frame  with an initial offset $(x_{\textrm{off}},z_{\textrm{off}})=(0.2a,0.2a)$.
Initially the system is not axisymmetric but after a slight rotation the swimmer settles in an axisymmetric configuration.   Although the rotational motion is small, it occurs early  in the dynamics, in particular before the swimmer closely approaches the droplet.  After that, the system becomes equivalent to the axisymmetric situation considered in Fig~\ref{fig:stability}c  and the swimmer reaches a stable state maintaining a thin gap with the droplet.

Next we show in Fig.~\ref{fig:tra_off}b  the trajectories of pushers with an initial offset $(x_{\textrm{off}},z_{\textrm{off}})=(0.2a,-0.2a)$.
The swimmer slightly rotates but in this case does not align with the droplet axisymmetrically.
Instead, due to the attractive flows  in the lateral directions,
the pusher approaches the droplet and eventually  collides with it.
Other cases with the initial offset $(x_{\textrm{off}},z_{\textrm{off}})=(0.2a,0.2a)$ or $(0.2a,0)$ 
exhibit similar behaviors as in Fig.~\ref{fig:tra_off}b with no stable configurations.
Also pullers with the initial offset $(x_{\textrm{off}},z_{\textrm{off}})=(0.2a,-0.2a)$ or $(0.2a,0)$ 
and neutral swimmers with $(x_{\textrm{off}},z_{\textrm{off}})=(0.2a,\pm0.2a)$ or $(0.2a,0)$ 
do not settle a stable configuration.
Additional simulations by changing the  size ratio and stresslet strength  leads to similar results.

\section{Conclusion}

In this paper, we  have studied in the creeping flow regime the dynamics of
a spherical squirmer encapsulated in an undeformable 
droplet using both theory and computations. The incompressible Stokes
equations were first solved analytically, and when the swimmer and droplet are 
concentric,
we obtained exact solutions of the \rc{}
swimmer and droplet velocities, the flow velocity fields and 
its dissipated power.
Along with this analytic approach, numerical simulations based on a 
boundary element method were performed
and the numerical results agreed well with the theoretical results.

The analytical solutions provide a useful physical picture of the instantaneous dynamics
for the concentric configuration of the squirmer and droplet. 
For a squirmer using pure tangential surface actuations, although their movement are doomed to be transient, 
the theoretical results state that the swimmer is always faster than the droplet.
When the normal surface velocities are incorporated on top of tangential modes,
the squirmer and droplet are able to co-swim with a same velocity and thus to remain concentric.

When the swimmers are slightly displaced from the concentric position,
we found that they would either return to the center (stable), deviate further and eventually touch the droplet interface (unstable), or reach an eccentric steady-state position (stable).
Such final states depend on swimming gaits or relative locations of swimmers.

The ultimate goal of encaging swimmers is to help  
transport and deliver small chemical  payloads, and thus a lot of future work 
lies ahead for swimmer-droplet complexes. Questions including swimming near 
complex boundaries or near walls, or non-axisymmetrically, will have to be 
tackled. Surfactants, which  are commonly used in droplet-based 
microfluidics to  prevent coalescence, could perhaps be used here
to prevent collision between swimmers and interface, with interesting physical 
consequences. Finally, if heterogeneous fluid mixtures are to be transported in 
the droplet, it will important to quantify their mixing and chemical fate as 
they move along with the 
swimmer.

\section{Acknowledgements}
Gioele Balestra is acknowledged for his helpful suggestions on making the $3$D 
schematic plot.
The computer time is provided by the Swiss National 
Supercomputing Centre (CSCS) under project ID s603 and by SNIC (Swedish 
National Infrastructure for Computing). A VR International Postdoc 
Grant from Swedish Research Council (L.Z.), an ERC starting grant 'SimCoMiCs 
280117' (F.G.), a Marie Curie CIG Grant (E.L.) and an ERC Consolidator grant (E.L.) are gratefully 
acknowledged.

\appendix

\section{Constants in flow solution}
The undefined constants for the fluid velocity fields in Eq.~\ref{vel_full} 
and the power calculations in Eq.~\ref{pw} are given  in 
Table  \ref{table1}.

\begin{table}
  \centering
  \caption{\label{table} The constants for the fluid velocity field given in Eq.~\ref{vel_full}
    and the power in Eq.~\ref{pw}.
  }
  \begin{tabular*}{0.95\textwidth}{@{\extracolsep{\fill}}c}
    \hline
    \vbox{
      \begin{align}
        &\Delta_n = \lambda\{(2n+1)^2(\chi^{2n-1}-1)(\chi^{2n+3}-1) \nonumber \\
        &\hspace{35pt}
        -(2n-1)(2n+3)(\chi^{2n+1}-1)^2\} \hspace{10pt}(n\geq 2)\nonumber\\
        &N_1 = n(2n-1)\{(\chi^{2n+1}-1)\lambda+1\} \nonumber \\
        &\hspace{22pt}-(n-2)\{(2n+1)(\chi^{2n-1}-1)\lambda-2\chi^{2n-1}+2n+1\}\nonumber\\
        &N_2 = -2(2n-1)\{(\chi^{2n+1}-1)\lambda+1\}\nonumber\\
        &\hspace{22pt}+2\{(2n+1)(\chi^{2n-1}-1)\lambda-2\chi^{2n-1}+2n+1\}\nonumber\\
        &N_3 = (n-2)(2n+3)\{(\chi^{2n+1}-1)\lambda+1\}\nonumber\\
        &\hspace{22pt}-n\{(2n+1)(\chi^{2n+3}-1)\lambda+2\chi^{2n+3}+2n+1\}\nonumber\\
        &N_4 = -2(2n+3)\{(\chi^{2n+1}-1)\lambda+1\}\nonumber\\
        &\hspace{22pt}+2\{(2n+1)(\chi^{2n+3}-1)\lambda+2\chi^{2n+3}+2n+1\},\nonumber\\
        &N_5 = \chi^{4n+2}[-(n+1)(2n+3)\{(1-\chi^{-2n-1})\lambda+1\}\nonumber\\
        &\hspace{22pt}+(n+3)\{(2n+1)(1-\chi^{-2n-3})\lambda+2\chi^{-2n-3}+2n+1\}]\nonumber\\
        &N_6 = \chi^{4n+2}[-2(2n+3)\{(1-\chi^{-2n-1})\lambda+1\}\nonumber\\
        &\hspace{22pt}+2\{(2n+1)(1-\chi^{-2n-3})\lambda+2\chi^{-2n-3}+2n+1\}]\nonumber\\
        &N_7 = \chi^{4n+2}[-(n+3)(2n-1)\{(1-\chi^{-2n-1})\lambda+1\}\nonumber\\
        &\hspace{22pt}+(n+1)\{(2n+1)(1-\chi^{-2n+1})\lambda-2\chi^{-2n+1}+2n+1\}]\nonumber\\
        &N_8 = \chi^{4n+2}[-2(2n-1)\{(1-\chi^{-2n-1})\lambda+1\}\nonumber\\
        &\hspace{22pt}+2\{(2n+1)(1-\chi^{-2n+1})\lambda-2\chi^{-2n+1}+2n+1\}]\nonumber\\
        &c_1 = -\frac{n+3}{2}N_1\chi^{2n+3}-\frac{n+1}{2}N_3\chi^{2n+1}+\frac{n-2}{2}N_5\chi^{2}+\frac{n}{2}N_7\nonumber\\
        &c_2 = -\frac{n+3}{2}N_2\chi^{2n+3}-\frac{n+1}{2}N_4\chi^{2n+1}+\frac{n-2}{2}N_6\chi^{2}+\frac{n}{2}N_8\nonumber \\
        &a_n = \bigg(\frac{2n+3}{n}-n-1 \bigg)N_1 - (n-1)N_3 
        +\bigg(n+\frac{2n-1}{n+1}\bigg)N_5 
        +(n+2)N_7 \nonumber\\
        &b_n = \bigg(\frac{2n+3}{n}-n-1 \bigg)N_2 - (n-1)N_4
        +\bigg(n+\frac{2n-1}{n+1}\bigg)N_6 
        +(n+2)N_8 \nonumber\\
        &c_n = n(n+2)N_1 + (n-1)(n+1)N_3 + (n-1)(n+1)N_5  
        +n(n+2)N_7 \nonumber\\
        &d_n = n(n+2)N_2 + (n-1)(n+1)N_4 + (n-1)(n+1)N_6 
        + n(n+2)N_8 \nonumber\\
        &N_o = N_1 + N_3 + N_5 + N_7 \nonumber \\
        &N_e = N_2 + N_4 + N_6 + N_8 \nonumber \\
        &\bar{N}_o = -\frac{n+3}{2}N_1 - \frac{n+1}{2}N_3 + \frac{n-2}{2}N_5 + \frac{n}{2}N_7 \nonumber \\
        &\bar{N}_e = -\frac{n+3}{2}N_2 - \frac{n+1}{2}N_4 + \frac{n-2}{2}N_6 + \frac{n}{2}N_8 \nonumber\\
        &Z_1 = 2(2\lambda+3)\chi^5-10(\lambda-1)\chi^2+6(\lambda-1) \nonumber\\
        &Z_2 = (2\lambda+3)\chi^5+10(\lambda-1)\chi^2-12(\lambda-1) \nonumber\\
        &Z_3 = 2\{(2\lambda+3)\chi^5+3(\lambda-1)\} \nonumber
      \end{align}
    }\\
    \hline
  \end{tabular*}
  \label{table1}
\end{table}

\newpage


\begin{thebibliography}{41}%
\makeatletter
\providecommand \@ifxundefined [1]{%
 \@ifx{#1\undefined}
}%
\providecommand \@ifnum [1]{%
 \ifnum #1\expandafter \@firstoftwo
 \else \expandafter \@secondoftwo
 \fi
}%
\providecommand \@ifx [1]{%
 \ifx #1\expandafter \@firstoftwo
 \else \expandafter \@secondoftwo
 \fi
}%
\providecommand \natexlab [1]{#1}%
\providecommand \enquote  [1]{``#1''}%
\providecommand \bibnamefont  [1]{#1}%
\providecommand \bibfnamefont [1]{#1}%
\providecommand \citenamefont [1]{#1}%
\providecommand \href@noop [0]{\@secondoftwo}%
\providecommand \href [0]{\begingroup \@sanitize@url \@href}%
\providecommand \@href[1]{\@@startlink{#1}\@@href}%
\providecommand \@@href[1]{\endgroup#1\@@endlink}%
\providecommand \@sanitize@url [0]{\catcode `\\12\catcode `\$12\catcode
  `\&12\catcode `\#12\catcode `\^12\catcode `\_12\catcode `\%12\relax}%
\providecommand \@@startlink[1]{}%
\providecommand \@@endlink[0]{}%
\providecommand \url  [0]{\begingroup\@sanitize@url \@url }%
\providecommand \@url [1]{\endgroup\@href {#1}{\urlprefix }}%
\providecommand \urlprefix  [0]{URL }%
\providecommand \Eprint [0]{\href }%
\providecommand \doibase [0]{http://dx.doi.org/}%
\providecommand \selectlanguage [0]{\@gobble}%
\providecommand \bibinfo  [0]{\@secondoftwo}%
\providecommand \bibfield  [0]{\@secondoftwo}%
\providecommand \translation [1]{[#1]}%
\providecommand \BibitemOpen [0]{}%
\providecommand \bibitemStop [0]{}%
\providecommand \bibitemNoStop [0]{.\EOS\space}%
\providecommand \EOS [0]{\spacefactor3000\relax}%
\providecommand \BibitemShut  [1]{\csname bibitem#1\endcsname}%
\let\auto@bib@innerbib\@empty
\bibitem [{\citenamefont {He}\ \emph {et~al.}(2005)\citenamefont {He},
  \citenamefont {Edgar}, \citenamefont {Jeffries}, \citenamefont {Lorenz},
  \citenamefont {Shelby},\ and\ \citenamefont {Chiu}}]{he2005selective}%
  \BibitemOpen
  \bibfield  {author} {\bibinfo {author} {\bibfnamefont {M.}~\bibnamefont
  {He}}, \bibinfo {author} {\bibfnamefont {J.~S.}\ \bibnamefont {Edgar}},
  \bibinfo {author} {\bibfnamefont {G.~D.}\ \bibnamefont {Jeffries}}, \bibinfo
  {author} {\bibfnamefont {R.~M.}\ \bibnamefont {Lorenz}}, \bibinfo {author}
  {\bibfnamefont {J.~P.}\ \bibnamefont {Shelby}}, \ and\ \bibinfo {author}
  {\bibfnamefont {D.~T.}\ \bibnamefont {Chiu}},\ }\href@noop {} {\bibfield
  {journal} {\bibinfo  {journal} {Anal. Chem.}\ }\textbf {\bibinfo {volume}
  {77}},\ \bibinfo {pages} {1539} (\bibinfo {year} {2005})}\BibitemShut
  {NoStop}%
\bibitem [{\citenamefont {K{\"o}ster}\ \emph {et~al.}(2008)\citenamefont
  {K{\"o}ster}, \citenamefont {Angile}, \citenamefont {Duan}, \citenamefont
  {Agresti}, \citenamefont {Wintner}, \citenamefont {Schmitz}, \citenamefont
  {Rowat}, \citenamefont {Merten}, \citenamefont {Pisignano}, \citenamefont
  {Griffiths} \emph {et~al.}}]{koster2008drop}%
  \BibitemOpen
  \bibfield  {author} {\bibinfo {author} {\bibfnamefont {S.}~\bibnamefont
  {K{\"o}ster}}, \bibinfo {author} {\bibfnamefont {F.~E.}\ \bibnamefont
  {Angile}}, \bibinfo {author} {\bibfnamefont {H.}~\bibnamefont {Duan}},
  \bibinfo {author} {\bibfnamefont {J.~J.}\ \bibnamefont {Agresti}}, \bibinfo
  {author} {\bibfnamefont {A.}~\bibnamefont {Wintner}}, \bibinfo {author}
  {\bibfnamefont {C.}~\bibnamefont {Schmitz}}, \bibinfo {author} {\bibfnamefont
  {A.~C.}\ \bibnamefont {Rowat}}, \bibinfo {author} {\bibfnamefont {C.~A.}\
  \bibnamefont {Merten}}, \bibinfo {author} {\bibfnamefont {D.}~\bibnamefont
  {Pisignano}}, \bibinfo {author} {\bibfnamefont {A.~D.}\ \bibnamefont
  {Griffiths}},  \emph {et~al.},\ }\href@noop {} {\bibfield  {journal}
  {\bibinfo  {journal} {Lab Chip}\ }\textbf {\bibinfo {volume} {8}},\ \bibinfo
  {pages} {1110} (\bibinfo {year} {2008})}\BibitemShut {NoStop}%
\bibitem [{\citenamefont {Chabert}\ and\ \citenamefont
  {Viovy}(2008)}]{chabert2008microfluidic}%
  \BibitemOpen
  \bibfield  {author} {\bibinfo {author} {\bibfnamefont {M.}~\bibnamefont
  {Chabert}}\ and\ \bibinfo {author} {\bibfnamefont {J.-L.}\ \bibnamefont
  {Viovy}},\ }\href@noop {} {\bibfield  {journal} {\bibinfo  {journal} {Proc.
  Natl. Acad. Sci. U.S.A.}\ }\textbf {\bibinfo {volume} {105}},\ \bibinfo
  {pages} {3191} (\bibinfo {year} {2008})}\BibitemShut {NoStop}%
\bibitem [{\citenamefont {Clausell-Tormos}\ \emph {et~al.}(2008)\citenamefont
  {Clausell-Tormos}, \citenamefont {Lieber}, \citenamefont {Baret},
  \citenamefont {El-Harrak}, \citenamefont {Miller}, \citenamefont {Frenz},
  \citenamefont {Blouwolff}, \citenamefont {Humphry}, \citenamefont
  {K{\"o}ster}, \citenamefont {Duan} \emph {et~al.}}]{clausell2008droplet}%
  \BibitemOpen
  \bibfield  {author} {\bibinfo {author} {\bibfnamefont {J.}~\bibnamefont
  {Clausell-Tormos}}, \bibinfo {author} {\bibfnamefont {D.}~\bibnamefont
  {Lieber}}, \bibinfo {author} {\bibfnamefont {J.-C.}\ \bibnamefont {Baret}},
  \bibinfo {author} {\bibfnamefont {A.}~\bibnamefont {El-Harrak}}, \bibinfo
  {author} {\bibfnamefont {O.~J.}\ \bibnamefont {Miller}}, \bibinfo {author}
  {\bibfnamefont {L.}~\bibnamefont {Frenz}}, \bibinfo {author} {\bibfnamefont
  {J.}~\bibnamefont {Blouwolff}}, \bibinfo {author} {\bibfnamefont {K.~J.}\
  \bibnamefont {Humphry}}, \bibinfo {author} {\bibfnamefont {S.}~\bibnamefont
  {K{\"o}ster}}, \bibinfo {author} {\bibfnamefont {H.}~\bibnamefont {Duan}},
  \emph {et~al.},\ }\href@noop {} {\bibfield  {journal} {\bibinfo  {journal}
  {Chem. Biol.}\ }\textbf {\bibinfo {volume} {15}},\ \bibinfo {pages} {427}
  (\bibinfo {year} {2008})}\BibitemShut {NoStop}%
\bibitem [{\citenamefont {Wen}\ \emph {et~al.}(2015)\citenamefont {Wen},
  \citenamefont {Yu}, \citenamefont {Zhu}, \citenamefont {Jiang},\ and\
  \citenamefont {Qin}}]{wen2015droplet}%
  \BibitemOpen
  \bibfield  {author} {\bibinfo {author} {\bibfnamefont {H.}~\bibnamefont
  {Wen}}, \bibinfo {author} {\bibfnamefont {Y.}~\bibnamefont {Yu}}, \bibinfo
  {author} {\bibfnamefont {G.}~\bibnamefont {Zhu}}, \bibinfo {author}
  {\bibfnamefont {L.}~\bibnamefont {Jiang}}, \ and\ \bibinfo {author}
  {\bibfnamefont {J.}~\bibnamefont {Qin}},\ }\href@noop {} {\bibfield
  {journal} {\bibinfo  {journal} {Lab Chip}\ }\textbf {\bibinfo {volume}
  {15}},\ \bibinfo {pages} {1905} (\bibinfo {year} {2015})}\BibitemShut
  {NoStop}%
\bibitem [{\citenamefont {Zhang}\ \emph {et~al.}(2009)\citenamefont {Zhang},
  \citenamefont {Abbott}, \citenamefont {Dong}, \citenamefont {Kratochvil},
  \citenamefont {Bell},\ and\ \citenamefont {Nelson}}]{zhang2009artificial}%
  \BibitemOpen
  \bibfield  {author} {\bibinfo {author} {\bibfnamefont {L.}~\bibnamefont
  {Zhang}}, \bibinfo {author} {\bibfnamefont {J.~J.}\ \bibnamefont {Abbott}},
  \bibinfo {author} {\bibfnamefont {L.}~\bibnamefont {Dong}}, \bibinfo {author}
  {\bibfnamefont {B.~E.}\ \bibnamefont {Kratochvil}}, \bibinfo {author}
  {\bibfnamefont {D.}~\bibnamefont {Bell}}, \ and\ \bibinfo {author}
  {\bibfnamefont {B.~J.}\ \bibnamefont {Nelson}},\ }\href@noop {} {\bibfield
  {journal} {\bibinfo  {journal} {Appl. Phys. Lett.}\ }\textbf {\bibinfo
  {volume} {94}},\ \bibinfo {pages} {064107} (\bibinfo {year}
  {2009})}\BibitemShut {NoStop}%
\bibitem [{\citenamefont {Tottori}\ \emph {et~al.}(2012)\citenamefont
  {Tottori}, \citenamefont {Zhang}, \citenamefont {Qiu}, \citenamefont
  {Krawczyk}, \citenamefont {Franco-Obreg{\'o}n},\ and\ \citenamefont
  {Nelson}}]{tottori2012magnetic}%
  \BibitemOpen
  \bibfield  {author} {\bibinfo {author} {\bibfnamefont {S.}~\bibnamefont
  {Tottori}}, \bibinfo {author} {\bibfnamefont {L.}~\bibnamefont {Zhang}},
  \bibinfo {author} {\bibfnamefont {F.}~\bibnamefont {Qiu}}, \bibinfo {author}
  {\bibfnamefont {K.~K.}\ \bibnamefont {Krawczyk}}, \bibinfo {author}
  {\bibfnamefont {A.}~\bibnamefont {Franco-Obreg{\'o}n}}, \ and\ \bibinfo
  {author} {\bibfnamefont {B.~J.}\ \bibnamefont {Nelson}},\ }\href@noop {}
  {\bibfield  {journal} {\bibinfo  {journal} {Adv. Mater.}\ }\textbf {\bibinfo
  {volume} {24}},\ \bibinfo {pages} {811} (\bibinfo {year} {2012})}\BibitemShut
  {NoStop}%
\bibitem [{\citenamefont {Ding}\ \emph {et~al.}(2016)\citenamefont {Ding},
  \citenamefont {Qiu}, \citenamefont {Solvas}, \citenamefont {Chiu},
  \citenamefont {Nelson},\ and\ \citenamefont {deMello}}]{ding:16}%
  \BibitemOpen
  \bibfield  {author} {\bibinfo {author} {\bibfnamefont {Y.}~\bibnamefont
  {Ding}}, \bibinfo {author} {\bibfnamefont {F.}~\bibnamefont {Qiu}}, \bibinfo
  {author} {\bibfnamefont {X.~C.}\ \bibnamefont {Solvas}}, \bibinfo {author}
  {\bibfnamefont {F.~W.~Y.}\ \bibnamefont {Chiu}}, \bibinfo {author}
  {\bibfnamefont {B.~J.}\ \bibnamefont {Nelson}}, \ and\ \bibinfo {author}
  {\bibfnamefont {A.}~\bibnamefont {deMello}},\ }\href@noop {} {\bibfield
  {journal} {\bibinfo  {journal} {Micromachines}\ }\textbf {\bibinfo {volume}
  {7}},\ \bibinfo {pages} {25} (\bibinfo {year} {2016})}\BibitemShut {NoStop}%
\bibitem [{\citenamefont {Lighthill}(1952)}]{lighthill:52}%
  \BibitemOpen
  \bibfield  {author} {\bibinfo {author} {\bibfnamefont {M.~J.}\ \bibnamefont
  {Lighthill}},\ }\href@noop {} {\bibfield  {journal} {\bibinfo  {journal}
  {Comm. Pure Appl. Math.}\ }\textbf {\bibinfo {volume} {5}},\ \bibinfo {pages}
  {109} (\bibinfo {year} {1952})}\BibitemShut {NoStop}%
\bibitem [{\citenamefont {Blake}(1971)}]{blake:71}%
  \BibitemOpen
  \bibfield  {author} {\bibinfo {author} {\bibfnamefont {J.~R.}\ \bibnamefont
  {Blake}},\ }\href@noop {} {\bibfield  {journal} {\bibinfo  {journal} {J.
  Fluid Mech.}\ }\textbf {\bibinfo {volume} {46}},\ \bibinfo {pages} {199}
  (\bibinfo {year} {1971})}\BibitemShut {NoStop}%
\bibitem [{\citenamefont {Magar}\ \emph {et~al.}(2003)\citenamefont {Magar},
  \citenamefont {Goto},\ and\ \citenamefont {Pedley}}]{magar2003nutrient}%
  \BibitemOpen
  \bibfield  {author} {\bibinfo {author} {\bibfnamefont {V.}~\bibnamefont
  {Magar}}, \bibinfo {author} {\bibfnamefont {T.}~\bibnamefont {Goto}}, \ and\
  \bibinfo {author} {\bibfnamefont {T.}~\bibnamefont {Pedley}},\ }\href@noop {}
  {\bibfield  {journal} {\bibinfo  {journal} {Q. J. Mech. Appl. Math.}\
  }\textbf {\bibinfo {volume} {56}},\ \bibinfo {pages} {65} (\bibinfo {year}
  {2003})}\BibitemShut {NoStop}%
\bibitem [{\citenamefont {Ishikawa}\ \emph {et~al.}(2006)\citenamefont
  {Ishikawa}, \citenamefont {Simmonds},\ and\ \citenamefont
  {Pedley}}]{ishikawa:06}%
  \BibitemOpen
  \bibfield  {author} {\bibinfo {author} {\bibfnamefont {T.}~\bibnamefont
  {Ishikawa}}, \bibinfo {author} {\bibfnamefont {M.~P.}\ \bibnamefont
  {Simmonds}}, \ and\ \bibinfo {author} {\bibfnamefont {T.~J.}\ \bibnamefont
  {Pedley}},\ }\href@noop {} {\bibfield  {journal} {\bibinfo  {journal} {J.
  Fluid Mech.}\ }\textbf {\bibinfo {volume} {568}},\ \bibinfo {pages} {119}
  (\bibinfo {year} {2006})}\BibitemShut {NoStop}%
\bibitem [{\citenamefont {Michelin}\ and\ \citenamefont
  {Lauga}(2010)}]{michelin2010efficiency}%
  \BibitemOpen
  \bibfield  {author} {\bibinfo {author} {\bibfnamefont {S.}~\bibnamefont
  {Michelin}}\ and\ \bibinfo {author} {\bibfnamefont {E.}~\bibnamefont
  {Lauga}},\ }\href@noop {} {\bibfield  {journal} {\bibinfo  {journal} {Phys.
  Fluids}\ }\textbf {\bibinfo {volume} {22}},\ \bibinfo {pages} {111901}
  (\bibinfo {year} {2010})}\BibitemShut {NoStop}%
\bibitem [{\citenamefont {Doostmohammadi}\ \emph {et~al.}(2012)\citenamefont
  {Doostmohammadi}, \citenamefont {Stocker},\ and\ \citenamefont
  {Ardekani}}]{doostmohammadi2012low}%
  \BibitemOpen
  \bibfield  {author} {\bibinfo {author} {\bibfnamefont {A.}~\bibnamefont
  {Doostmohammadi}}, \bibinfo {author} {\bibfnamefont {R.}~\bibnamefont
  {Stocker}}, \ and\ \bibinfo {author} {\bibfnamefont {A.~M.}\ \bibnamefont
  {Ardekani}},\ }\href@noop {} {\bibfield  {journal} {\bibinfo  {journal}
  {Proc. Natl. Acad. Sci. U.S.A.}\ }\textbf {\bibinfo {volume} {109}},\
  \bibinfo {pages} {3856} (\bibinfo {year} {2012})}\BibitemShut {NoStop}%
\bibitem [{\citenamefont {Z{\"o}ttl}\ and\ \citenamefont
  {Stark}(2012)}]{zottl2012nonlinear}%
  \BibitemOpen
  \bibfield  {author} {\bibinfo {author} {\bibfnamefont {A.}~\bibnamefont
  {Z{\"o}ttl}}\ and\ \bibinfo {author} {\bibfnamefont {H.}~\bibnamefont
  {Stark}},\ }\href@noop {} {\bibfield  {journal} {\bibinfo  {journal} {Phys.
  Rev. Lett.}\ }\textbf {\bibinfo {volume} {108}},\ \bibinfo {pages} {218104}
  (\bibinfo {year} {2012})}\BibitemShut {NoStop}%
\bibitem [{\citenamefont {Pak}\ and\ \citenamefont {Lauga}(2014)}]{pak:14}%
  \BibitemOpen
  \bibfield  {author} {\bibinfo {author} {\bibfnamefont {O.~S.}\ \bibnamefont
  {Pak}}\ and\ \bibinfo {author} {\bibfnamefont {E.}~\bibnamefont {Lauga}},\
  }\href@noop {} {\bibfield  {journal} {\bibinfo  {journal} {J. Eng. Math.}\
  }\textbf {\bibinfo {volume} {88}},\ \bibinfo {pages} {1} (\bibinfo {year}
  {2014})}\BibitemShut {NoStop}%
\bibitem [{\citenamefont {Datt}\ \emph {et~al.}(2015)\citenamefont {Datt},
  \citenamefont {Zhu}, \citenamefont {Elfring},\ and\ \citenamefont
  {Pak}}]{datt2015squirming}%
  \BibitemOpen
  \bibfield  {author} {\bibinfo {author} {\bibfnamefont {C.}~\bibnamefont
  {Datt}}, \bibinfo {author} {\bibfnamefont {L.}~\bibnamefont {Zhu}}, \bibinfo
  {author} {\bibfnamefont {G.~J.}\ \bibnamefont {Elfring}}, \ and\ \bibinfo
  {author} {\bibfnamefont {O.~S.}\ \bibnamefont {Pak}},\ }\href@noop {}
  {\bibfield  {journal} {\bibinfo  {journal} {J. Fluid Mech}\ }\textbf
  {\bibinfo {volume} {784}},\ \bibinfo {pages} {R1} (\bibinfo {year}
  {2015})}\BibitemShut {NoStop}%
\bibitem [{\citenamefont {Delfau}\ \emph {et~al.}(2016)\citenamefont {Delfau},
  \citenamefont {Molina},\ and\ \citenamefont {Sano}}]{delfau2016collective}%
  \BibitemOpen
  \bibfield  {author} {\bibinfo {author} {\bibfnamefont {J.-B.}\ \bibnamefont
  {Delfau}}, \bibinfo {author} {\bibfnamefont {J.}~\bibnamefont {Molina}}, \
  and\ \bibinfo {author} {\bibfnamefont {M.}~\bibnamefont {Sano}},\ }\href@noop
  {} {\bibfield  {journal} {\bibinfo  {journal} {EPL}\ }\textbf {\bibinfo
  {volume} {114}},\ \bibinfo {pages} {24001} (\bibinfo {year}
  {2016})}\BibitemShut {NoStop}%
\bibitem [{\citenamefont {Lambs}(1932)}]{lamb}%
  \BibitemOpen
  \bibfield  {author} {\bibinfo {author} {\bibfnamefont {H.}~\bibnamefont
  {Lambs}},\ }\href@noop {} {\emph {\bibinfo {title} {Hydrodynamics}}},\
  \bibinfo {edition} {6th}\ ed.\ (\bibinfo  {publisher} {Cambridge University
  Press},\ \bibinfo {year} {1932})\BibitemShut {NoStop}%
\bibitem [{\citenamefont {Happel}\ and\ \citenamefont
  {Brenner}(1973)}]{happel:73}%
  \BibitemOpen
  \bibfield  {author} {\bibinfo {author} {\bibfnamefont {J.}~\bibnamefont
  {Happel}}\ and\ \bibinfo {author} {\bibfnamefont {H.}~\bibnamefont
  {Brenner}},\ }\href@noop {} {\emph {\bibinfo {title} {Low Reynolds Number
  Hydrodynamics}}}\ (\bibinfo  {publisher} {Noordhoff International
  publishing},\ \bibinfo {address} {Leyden},\ \bibinfo {year}
  {1973})\BibitemShut {NoStop}%
\bibitem [{\citenamefont {Pozrikidis}(1992)}]{pozrikidis1992boundary}%
  \BibitemOpen
  \bibfield  {author} {\bibinfo {author} {\bibfnamefont {C.}~\bibnamefont
  {Pozrikidis}},\ }\href@noop {} {\emph {\bibinfo {title} {Boundary integral
  and singularity methods for linearized viscous flow}}}\ (\bibinfo
  {publisher} {Cambridge University Press},\ \bibinfo {year}
  {1992})\BibitemShut {NoStop}%
\bibitem [{\citenamefont {Higdon}\ and\ \citenamefont
  {Muldowney}(1995)}]{tube_higdon}%
  \BibitemOpen
  \bibfield  {author} {\bibinfo {author} {\bibfnamefont {J.~J.~L.}\
  \bibnamefont {Higdon}}\ and\ \bibinfo {author} {\bibfnamefont {G.~P.}\
  \bibnamefont {Muldowney}},\ }\href@noop {} {\bibfield  {journal} {\bibinfo
  {journal} {J. Fluid Mech.}\ }\textbf {\bibinfo {volume} {298}},\ \bibinfo
  {pages} {193} (\bibinfo {year} {1995})}\BibitemShut {NoStop}%
\bibitem [{\citenamefont {Zhu}\ \emph {et~al.}(2013)\citenamefont {Zhu},
  \citenamefont {Lauga},\ and\ \citenamefont {Brandt}}]{zhu2013low}%
  \BibitemOpen
  \bibfield  {author} {\bibinfo {author} {\bibfnamefont {L.}~\bibnamefont
  {Zhu}}, \bibinfo {author} {\bibfnamefont {E.}~\bibnamefont {Lauga}}, \ and\
  \bibinfo {author} {\bibfnamefont {L.}~\bibnamefont {Brandt}},\ }\href@noop {}
  {\bibfield  {journal} {\bibinfo  {journal} {J. Fluid Mech.}\ }\textbf
  {\bibinfo {volume} {726}},\ \bibinfo {pages} {285} (\bibinfo {year}
  {2013})}\BibitemShut {NoStop}%
\bibitem [{\citenamefont {Dunavant}(1985)}]{dunavant1985high}%
  \BibitemOpen
  \bibfield  {author} {\bibinfo {author} {\bibfnamefont {D.}~\bibnamefont
  {Dunavant}},\ }\href@noop {} {\bibfield  {journal} {\bibinfo  {journal}
  {International journal for numerical methods in engineering}\ }\textbf
  {\bibinfo {volume} {21}},\ \bibinfo {pages} {1129} (\bibinfo {year}
  {1985})}\BibitemShut {NoStop}%
\bibitem [{\citenamefont {Pozrikidis}(2002)}]{poz_blue}%
  \BibitemOpen
  \bibfield  {author} {\bibinfo {author} {\bibfnamefont {C.}~\bibnamefont
  {Pozrikidis}},\ }\href@noop {} {\emph {\bibinfo {title} {A practical guide to
  boundary element methods with the software library BEMLIB}}},\ \bibinfo
  {edition} {1st}\ ed.\ (\bibinfo  {publisher} {CRC Press},\ \bibinfo {year}
  {2002})\BibitemShut {NoStop}%
\bibitem [{\citenamefont {Zinchenko}\ and\ \citenamefont
  {Davis}(2006)}]{zinchenko2006boundary}%
  \BibitemOpen
  \bibfield  {author} {\bibinfo {author} {\bibfnamefont {A.}~\bibnamefont
  {Zinchenko}}\ and\ \bibinfo {author} {\bibfnamefont {R.}~\bibnamefont
  {Davis}},\ }\href@noop {} {\bibfield  {journal} {\bibinfo  {journal} {J.
  Fluid Mech.}\ }\textbf {\bibinfo {volume} {564}},\ \bibinfo {pages} {227}
  (\bibinfo {year} {2006})}\BibitemShut {NoStop}%
\bibitem [{\citenamefont {Zinchenko}\ \emph {et~al.}(1997)\citenamefont
  {Zinchenko}, \citenamefont {Rother},\ and\ \citenamefont
  {Davis}}]{zinchenko1997novel}%
  \BibitemOpen
  \bibfield  {author} {\bibinfo {author} {\bibfnamefont {A.~Z.}\ \bibnamefont
  {Zinchenko}}, \bibinfo {author} {\bibfnamefont {M.~A.}\ \bibnamefont
  {Rother}}, \ and\ \bibinfo {author} {\bibfnamefont {R.~H.}\ \bibnamefont
  {Davis}},\ }\href@noop {} {\bibfield  {journal} {\bibinfo  {journal} {Phys.
  Fluids}\ }\textbf {\bibinfo {volume} {9}},\ \bibinfo {pages} {1493} (\bibinfo
  {year} {1997})}\BibitemShut {NoStop}%
\bibitem [{\citenamefont {Zinchenko}\ and\ \citenamefont
  {Davis}(2013)}]{zinchenko2013emulsion}%
  \BibitemOpen
  \bibfield  {author} {\bibinfo {author} {\bibfnamefont {A.}~\bibnamefont
  {Zinchenko}}\ and\ \bibinfo {author} {\bibfnamefont {R.}~\bibnamefont
  {Davis}},\ }\href@noop {} {\bibfield  {journal} {\bibinfo  {journal} {J.
  Fluid Mech.}\ }\textbf {\bibinfo {volume} {725}},\ \bibinfo {pages} {611}
  (\bibinfo {year} {2013})}\BibitemShut {NoStop}%
\bibitem [{\citenamefont {Zhu}\ and\ \citenamefont
  {Gallaire}(2016)}]{zhu16pancake}%
  \BibitemOpen
  \bibfield  {author} {\bibinfo {author} {\bibfnamefont {L.}~\bibnamefont
  {Zhu}}\ and\ \bibinfo {author} {\bibfnamefont {F.}~\bibnamefont {Gallaire}},\
  }\href@noop {} {\bibfield  {journal} {\bibinfo  {journal} {J. Fluid Mech.}\
  }\textbf {\bibinfo {volume} {798}},\ \bibinfo {pages} {955} (\bibinfo {year}
  {2016})}\BibitemShut {NoStop}%
\bibitem [{\citenamefont {Berg}(2004)}]{berg04}%
  \BibitemOpen
  \bibfield  {author} {\bibinfo {author} {\bibfnamefont {H.~C.}\ \bibnamefont
  {Berg}},\ }\href@noop {} {\emph {\bibinfo {title} {E. coli in Motion}}}\
  (\bibinfo  {publisher} {Springer},\ \bibinfo {address} {New York},\ \bibinfo
  {year} {2004})\BibitemShut {NoStop}%
\bibitem [{\citenamefont {Hernandez-Ortiz}\ \emph {et~al.}(2009)\citenamefont
  {Hernandez-Ortiz}, \citenamefont {Underhill},\ and\ \citenamefont
  {Graham}}]{graham:09}%
  \BibitemOpen
  \bibfield  {author} {\bibinfo {author} {\bibfnamefont {J.~P.}\ \bibnamefont
  {Hernandez-Ortiz}}, \bibinfo {author} {\bibfnamefont {P.~T.}\ \bibnamefont
  {Underhill}}, \ and\ \bibinfo {author} {\bibfnamefont {M.~D.}\ \bibnamefont
  {Graham}},\ }\href@noop {} {\bibfield  {journal} {\bibinfo  {journal} {J.
  Phys. Condens. Matter}\ }\textbf {\bibinfo {volume} {21}},\ \bibinfo {pages}
  {204107} (\bibinfo {year} {2009})}\BibitemShut {NoStop}%
\bibitem [{\citenamefont {Spagnolie}\ and\ \citenamefont
  {Lauga}(2012)}]{spag:12}%
  \BibitemOpen
  \bibfield  {author} {\bibinfo {author} {\bibfnamefont {S.~E.}\ \bibnamefont
  {Spagnolie}}\ and\ \bibinfo {author} {\bibfnamefont {E.}~\bibnamefont
  {Lauga}},\ }\href@noop {} {\bibfield  {journal} {\bibinfo  {journal} {J.
  Fluid. Mech.}\ }\textbf {\bibinfo {volume} {700}},\ \bibinfo {pages} {105}
  (\bibinfo {year} {2012})}\BibitemShut {NoStop}%
\bibitem [{\citenamefont {Goldstein}(2015)}]{goldstein:15}%
  \BibitemOpen
  \bibfield  {author} {\bibinfo {author} {\bibfnamefont {R.~E.}\ \bibnamefont
  {Goldstein}},\ }\href@noop {} {\bibfield  {journal} {\bibinfo  {journal}
  {Annu. Rev. Fluid. Mech.}\ }\textbf {\bibinfo {volume} {47}},\ \bibinfo
  {pages} {343} (\bibinfo {year} {2015})}\BibitemShut {NoStop}%
\bibitem [{\citenamefont {Yoshinaga}\ \emph {et~al.}(2012)\citenamefont
  {Yoshinaga}, \citenamefont {Nagai}, \citenamefont {Sumino},\ and\
  \citenamefont {Kitahata}}]{yoshi:12}%
  \BibitemOpen
  \bibfield  {author} {\bibinfo {author} {\bibfnamefont {N.}~\bibnamefont
  {Yoshinaga}}, \bibinfo {author} {\bibfnamefont {K.~H.}\ \bibnamefont
  {Nagai}}, \bibinfo {author} {\bibfnamefont {Y.}~\bibnamefont {Sumino}}, \
  and\ \bibinfo {author} {\bibfnamefont {H.}~\bibnamefont {Kitahata}},\
  }\href@noop {} {\bibfield  {journal} {\bibinfo  {journal} {Phys. Rev. E}\
  }\textbf {\bibinfo {volume} {86}},\ \bibinfo {pages} {016108} (\bibinfo
  {year} {2012})}\BibitemShut {NoStop}%
\bibitem [{\citenamefont {Schmitt}\ and\ \citenamefont
  {Stark}(2013)}]{schmitt:13}%
  \BibitemOpen
  \bibfield  {author} {\bibinfo {author} {\bibfnamefont {M.}~\bibnamefont
  {Schmitt}}\ and\ \bibinfo {author} {\bibfnamefont {H.}~\bibnamefont
  {Stark}},\ }\href@noop {} {\bibfield  {journal} {\bibinfo  {journal} {EPL}\
  }\textbf {\bibinfo {volume} {101}},\ \bibinfo {pages} {44008} (\bibinfo
  {year} {2013})}\BibitemShut {NoStop}%
\bibitem [{\citenamefont {Herminghaus}\ \emph {et~al.}(2014)\citenamefont
  {Herminghaus}, \citenamefont {Maass}, \citenamefont {Kr{\"u}ger},
  \citenamefont {Thutupalli}, \citenamefont {Goehring},\ and\ \citenamefont
  {Bahr}}]{herm:14}%
  \BibitemOpen
  \bibfield  {author} {\bibinfo {author} {\bibfnamefont {S.}~\bibnamefont
  {Herminghaus}}, \bibinfo {author} {\bibfnamefont {C.~C.}\ \bibnamefont
  {Maass}}, \bibinfo {author} {\bibfnamefont {C.}~\bibnamefont {Kr{\"u}ger}},
  \bibinfo {author} {\bibfnamefont {S.}~\bibnamefont {Thutupalli}}, \bibinfo
  {author} {\bibfnamefont {L.}~\bibnamefont {Goehring}}, \ and\ \bibinfo
  {author} {\bibfnamefont {C.}~\bibnamefont {Bahr}},\ }\href@noop {} {\bibfield
   {journal} {\bibinfo  {journal} {Soft Matter}\ }\textbf {\bibinfo {volume}
  {10}},\ \bibinfo {pages} {7008} (\bibinfo {year} {2014})}\BibitemShut
  {NoStop}%
\bibitem [{\citenamefont {Maass}\ \emph {et~al.}(2016)\citenamefont {Maass},
  \citenamefont {Kr{\"u}ger}, \citenamefont {Herminghaus},\ and\ \citenamefont
  {Bahr}}]{herm:16}%
  \BibitemOpen
  \bibfield  {author} {\bibinfo {author} {\bibfnamefont {C.~C.}\ \bibnamefont
  {Maass}}, \bibinfo {author} {\bibfnamefont {C.}~\bibnamefont {Kr{\"u}ger}},
  \bibinfo {author} {\bibfnamefont {S.}~\bibnamefont {Herminghaus}}, \ and\
  \bibinfo {author} {\bibfnamefont {C.}~\bibnamefont {Bahr}},\ }\href@noop {}
  {\bibfield  {journal} {\bibinfo  {journal} {Anuu. Rev. Condens. Matter
  Phys.}\ }\textbf {\bibinfo {volume} {7}},\ \bibinfo {pages} {171} (\bibinfo
  {year} {2016})}\BibitemShut {NoStop}%
\bibitem [{\citenamefont {Golestanian}\ \emph {et~al.}(2005)\citenamefont
  {Golestanian}, \citenamefont {Liverpool},\ and\ \citenamefont
  {Ajdari}}]{goles:05}%
  \BibitemOpen
  \bibfield  {author} {\bibinfo {author} {\bibfnamefont {R.}~\bibnamefont
  {Golestanian}}, \bibinfo {author} {\bibfnamefont {T.~B.}\ \bibnamefont
  {Liverpool}}, \ and\ \bibinfo {author} {\bibfnamefont {A.}~\bibnamefont
  {Ajdari}},\ }\href@noop {} {\bibfield  {journal} {\bibinfo  {journal} {Phys.
  Rev. Lett.}\ }\textbf {\bibinfo {volume} {94}},\ \bibinfo {pages} {220801}
  (\bibinfo {year} {2005})}\BibitemShut {NoStop}%
\bibitem [{\citenamefont {Anderson}(1989)}]{anderson:89}%
  \BibitemOpen
  \bibfield  {author} {\bibinfo {author} {\bibfnamefont {J.~L.}\ \bibnamefont
  {Anderson}},\ }\href@noop {} {\bibfield  {journal} {\bibinfo  {journal}
  {Annu. Rev. Fluid. Mech.}\ }\textbf {\bibinfo {volume} {21}},\ \bibinfo
  {pages} {61} (\bibinfo {year} {1989})}\BibitemShut {NoStop}%
\bibitem [{\citenamefont {Wang}\ \emph {et~al.}(2013)\citenamefont {Wang},
  \citenamefont {Duan}, \citenamefont {Ahmed}, \citenamefont {Mallouk},\ and\
  \citenamefont {Sen}}]{wang:13}%
  \BibitemOpen
  \bibfield  {author} {\bibinfo {author} {\bibfnamefont {W.}~\bibnamefont
  {Wang}}, \bibinfo {author} {\bibfnamefont {W.}~\bibnamefont {Duan}}, \bibinfo
  {author} {\bibfnamefont {S.}~\bibnamefont {Ahmed}}, \bibinfo {author}
  {\bibfnamefont {T.~E.}\ \bibnamefont {Mallouk}}, \ and\ \bibinfo {author}
  {\bibfnamefont {A.}~\bibnamefont {Sen}},\ }\href@noop {} {\bibfield
  {journal} {\bibinfo  {journal} {Nano Today}\ }\textbf {\bibinfo {volume}
  {8}},\ \bibinfo {pages} {531} (\bibinfo {year} {2013})}\BibitemShut {NoStop}%
\bibitem [{\citenamefont {Colberg}\ \emph {et~al.}(2014)\citenamefont
  {Colberg}, \citenamefont {Reigh}, \citenamefont {Robertson},\ and\
  \citenamefont {Kapral}}]{colberg:14}%
  \BibitemOpen
  \bibfield  {author} {\bibinfo {author} {\bibfnamefont {P.~H.}\ \bibnamefont
  {Colberg}}, \bibinfo {author} {\bibfnamefont {S.~Y.}\ \bibnamefont {Reigh}},
  \bibinfo {author} {\bibfnamefont {B.}~\bibnamefont {Robertson}}, \ and\
  \bibinfo {author} {\bibfnamefont {R.}~\bibnamefont {Kapral}},\ }\href@noop {}
  {\bibfield  {journal} {\bibinfo  {journal} {Acc. Chem. Res.}\ }\textbf
  {\bibinfo {volume} {47}},\ \bibinfo {pages} {3504} (\bibinfo {year}
  {2014})}\BibitemShut {NoStop}%
\end{thebibliography}

%

\end{document}